\documentclass[12pt]{article}
\usepackage{amsfonts}

\usepackage{graphics}
\usepackage{amsmath}

\input{tcilatex}

\begin{document}

\centerline{\textbf{\Large}}

\vskip 0.8truecm

\centerline{\textbf{\Large BOUND-STATE VARIATIONAL WAVE EQUATION}} \vskip %
0.3truecm \centerline{\textbf{\Large FOR FERMION SYSTEMS IN QED}}

\vskip 0.6truecm 
\centerline{\large Andrei G. Terekidi$^{a }$, 
Jurij W. Darewych$^{b }$, Marko Horbatsch$^{c}$}

\vskip 0.5truecm 
\centerline{\footnotesize \emph{Department of Physics and
Astronomy, York University, Toronto, Ontario, M3J 1P3, Canada}} 
\centerline{\footnotesize \emph{$^{a }$terekidi@yorku.ca,
$^{b}$darewych@yorku.ca,$^{c}$marko@yorku.ca } }

\vskip 0.6truecm

%\maketitle

\noindent \textbf{Abstract} We present a formulation of the Hamiltonian
variational method for QED which enables the derivation of relativistic
few-fermion wave equation that can account, at least in principle, for
interactions to any order of the coupling constant.\ We derive a
relativistic two-fermion wave equation using this approach. The interaction
kernel of the equation is shown to be the generalized invariant $\widetilde{%
\mathcal{M}}\mathcal{\,}\ $matrix including all orders of Feynman diagrams.
The result is obtained rigorously from the underlying QFT for arbitrary mass
ratio of the two fermions. Our approach is based on three key points: a
reformulation of QED, the variational method, and adiabatic hypothesis. As
an application we calculate the one-loop contribution of radiative
corrections to the two-fermion binding energy for singlet states with
arbitrary principal quantum number $n$, and $\ell =J=0$\ . Our calculations
are carried out in the explicitly covariant Feynman gauge.

%\newpage
\vskip1.6truecm

\noindent\textbf{\large 1. Introduction} %\section{Introduction}

\vskip0.4truecm The description of two-particle states (particularly bound
states) in QFT such as quantum electrodynamics, including relativistic
effects, is an important problem. It is well known that in the
nonrelativistic limit this problem has been solved on the basis of the
Shr\"{o}dinger equation. In the relativistic case we have the Dirac
equation, which describes a one-fermion particle system only. However the
implementation of this equation for two-particle or multi-particle system
meets significant difficulties. It requires a relativistic manybody
approach, which is provided by QFT. However, the QFT implementation to
relativistic few-particle bound states is difficult. The usual method for
treating two-body bound states in QFT is by means of the Bethe-Salpeter
equation. It has a number of shortcomings, among them, a perturbative
treatment of interactions, which is unsuitable for strongly coupled system.
Other difficulties are the appearance of relative time coordinates, which
make it problematic to interpret the Bethe-Salpeter amplitude as a
traditional wave function. There are various ways of avoiding the problem,
such as, e.g., single-time reductions, but these are generally non-rigorous
approximations to the original problem in QFT.

Recently we applied the variational Hamiltonian formalism to the problem of
two fermions interacting through the electromagnetic field [1,2]. We derived
two-fermion relativistic wave equations, classified all bound states and
calculated the relativistic energy spectrum to fourth order in the fine
structure constant. The results were obtained for arbitrary mass ratio of
the particles. As a consequence, the fourth-order hyperfine splitting
formulas were derived for arbitrary quantum numbers and mass ratio. These
formulas are in agreement with experimental results for hydrogen and
muonium, as well as with previous calculations.

In the present paper we develop the variational Hamiltonian formalism for
two-fermion systems to allow for the inclusion of effects of higher order,
such as vertex corrections, vacuum polarization, and two-photon exchange.
This can be achieved by using the variational principle in combination with
the adiabatic hypothesis. The two-fermion wave equations are derived
rigorously from underlying QFT for arbitrary mass ratio. The solution of
these equations allows one to calculate energy corrections to all orders on
the basis of a two-body approach. The method presented here can be easily
generalized to three- and multi-particle systems. It is important to note
that the variational Hamiltonian formalism is applicable to strongly coupled
systems for which perturbation theory may be unreliable.

\vskip0.8truecm \noindent \textbf{\large 2. Reformulated QED} 
%\section{Introduction}

\vskip0.4truecm

It has been pointed out in previous publications that various models in
Quantum Field Theory (QFT), including QED, can be reformulated, with the
help of mediating-field Green's functions, into a form that is particularly
convenient for the investigation of bound-state problems and variational
calculations [3,4]. This approach was applied to the study of relativistic
two-body eigenstates in the scalar Yukawa (Wick-Cutkosky) theory [5,6,7]. We
shall implement such an approach to two-fermion states in QED in this paper.

The Lagrangian for two fermionic fields interacting electromagnetically is ($%
\hbar =c=1$){\normalsize 
\begin{eqnarray}
\mathcal{L} &=&\overline{\psi }(x)\left( i\gamma ^{\mu }\partial _{\mu
}-m_{1}-Q_{1}\gamma ^{\mu }A_{\mu }(x)\right) \psi (x)+\overline{\phi }%
(x)\left( i\gamma ^{\mu }\partial _{\mu }-m_{2}-Q_{2}\gamma ^{\mu }A_{\mu
}(x)\right) \phi (x)  \notag \\
&&\;\;\;\;\;\;\;\;\;\;\;\;\;-\frac{1}{4}\left( \partial _{\alpha }A_{\beta
}(x)-\partial _{\beta }A_{\alpha }(x)\right) \left( \partial ^{\alpha
}A^{\beta }(x)-\partial ^{\beta }A^{\alpha }(x)\right) .
\end{eqnarray}
}The corresponding Euler-Lagrange equations of motion are the coupled
Dirac-Maxwell equations, {\normalsize 
\begin{equation}
\left( i\gamma ^{\mu }\partial _{\mu }-m_{1}\right) \psi (x)=Q_{1}\gamma
^{\mu }A_{\mu }(x)\psi (x),
\end{equation}
} 
\begin{equation}
\left( i\gamma ^{\mu }\partial _{\mu }-m_{2}\right) \phi (x)=Q_{2}\gamma
^{\mu }A_{\mu }(x)\phi (x),
\end{equation}
and 
\begin{equation}
\partial _{\mu }\partial ^{\mu }A^{\nu }(x)-\partial ^{\nu }\partial _{\mu
}A^{\mu }(x)=j^{\nu }(x),
\end{equation}
where 
\begin{equation}
j^{\nu }(x)=Q_{1}\overline{\psi }(x)\gamma ^{\nu }\psi (x)+Q_{2}\overline{%
\phi }(x)\gamma ^{\nu }\phi (x).
\end{equation}
Equations (2)-(4) can be decoupled in part by using the well-known formal
solution [6-10] of the Maxwell equation (4), namely 
\begin{equation}
A_{\mu }(x)=A_{\mu }^{0}(x)+\int d^{4}x^{\prime }D_{\mu \nu }(x-x^{\prime
})j^{\nu }(x^{\prime }),
\end{equation}
where $D_{\mu \nu }(x-x^{\prime })$ is a Green's function (or photon
propagator in QFT terminology), defined by 
\begin{equation}
\partial _{\alpha }\partial ^{\alpha }D_{\mu \nu }(x-x^{\prime })-\partial
_{\mu }\partial ^{\alpha }D_{\alpha \nu }(x-x^{\prime })=g_{\mu \nu }\delta
^{4}(x-x^{\prime }),
\end{equation}
and $A_{\mu }^{0}(x)$ is a solution of the homogeneous (or ``free field'')
equation (4) with $j^{\mu }(x)=0.$

Equation (7) does not define the covariant Green's function $D_{\mu \nu
}(x-x^{\prime })$ uniquely. One can always add a solution of the homogeneous
equation (eq. (7) with $g_{\mu \nu }\rightarrow 0$). This allows for a
certain freedom in the choice of $D_{\mu \nu }$, as is discussed in standard
texts (e.g. ref. [8,9]). In practice, the solution of eq. (7), like that \
of eq. (4), requires a choice of gauge. However, we do not need to specify
one at this stage.

Substitution of the formal solution (6) into equations (2) and (3) yields
the ``partly reduced'' equations,

{\normalsize 
\begin{eqnarray}
\left( i\gamma ^{\mu }\partial _{\mu }-m_{1}\right) \psi (x) &=&Q_{1}\gamma
^{\mu }\left( A_{\mu }^{0}(x)+\int d^{4}x^{\prime }D_{\mu \nu }(x-x^{\prime
})j^{\nu }(x^{\prime })\right) \psi (x), \\
\left( i\gamma ^{\mu }\partial _{\mu }-m_{2}\right) \phi (x) &=&Q_{2}\gamma
^{\mu }\left( A_{\mu }^{0}(x)+\int d^{4}x^{\prime }D_{\mu \nu }(x-x^{\prime
})j^{\nu }(x^{\prime })\right) \phi (x).
\end{eqnarray}
}These are coupled nonlinear Dirac equations. To our knowledge no exact
(analytic or numeric) solution of equations (8)-(9) for classical fields
have been reported in the literature, though approximate (perturbative)
solutions have been discussed by various authors, particularly Barut and his
co-workers (see ref. [10,11] and citations therein). However, our interest
here is in the quantized field theory.

The partially reduced equations (8)-(9) are derivable from the stationary
action principle 
\begin{equation}
\delta S\left[ \psi ,\phi \right] =\delta \int d^{4}x\mathcal{L}_{R}=0
\end{equation}
with the Lagrangian density {\normalsize 
\begin{eqnarray}
\mathcal{L}_{R} &=&\overline{\psi }(x)\left( i\gamma ^{\mu }\partial _{\mu
}-m_{1}-Q_{1}\gamma ^{\mu }A_{\mu }^{0}(x)\right) \psi (x)+\overline{\phi }%
(x)\left( i\gamma ^{\mu }\partial _{\mu }-m_{2}-Q_{2}\gamma ^{\mu }A_{\mu
}^{0}(x)\right) \phi (x)  \notag \\
&&\;\;\;\;\;\;\;\;\;\;\;\;\;\;\;\;\;\;\;-\frac{1}{2}\int d^{4}x^{\prime
}j^{\mu }(x^{\prime })D_{\mu \nu }(x-x^{\prime })j^{\nu }(x)
\end{eqnarray}
}provided that the Green's function is symmetric in the sense that 
\begin{equation}
D_{\mu \nu }(x-x^{\prime })=D_{\mu \nu }(x^{\prime }-x),\;\;\;\;\;\;\;\;\;\
\;\ D_{\mu \nu }(x-x^{\prime })=D_{\nu \mu }(x-x^{\prime }).
\end{equation}
The interaction part of (11) has a somewhat modified structure from that of
the usual formulation of QED. Thus, there are two interaction terms. The
last term of (11) is a ``current-current'' interaction which contains the
photon propagator sandwiched between the fermionic currents. We shall use
this modified formulation together with a variational approach to obtain
relativistic two-fermion equations, and will study their bound-state
solutions.

We consider the quantized theory in the Hamiltonian equal-time formalism. To
this end we write down the Hamiltonian density corresponding to the
Lagrangian (11), namely 
\begin{equation}
\mathcal{H}_{R}=\mathcal{H}_{0}\mathcal{+H}_{I},\;\;\;\;\;\;\;\;\;\mathcal{H}%
_{I}=\mathcal{H}_{I_{1}}+\mathcal{H}_{I_{2}},
\end{equation}
where 
\begin{eqnarray}
\mathcal{H}_{0} &=&\psi ^{\dagger }(x)\left( -i\overrightarrow{\alpha }\cdot
\nabla +m_{1}\beta \right) \psi (x)+\phi ^{\dagger }(x)\left( -i%
\overrightarrow{\alpha }\cdot \nabla +m_{2}\beta \right) \phi (x), \\
\mathcal{H}_{I_{1}} &=&\frac{1}{2}\int d^{4}x^{\prime }j^{\mu }(x^{\prime
})D_{\mu \nu }(x-x^{\prime })j^{\nu }(x), \\
\mathcal{H}_{I_{2}} &=&Q_{1}\overline{\psi }(x)\,\gamma ^{\mu }A_{\mu
}^{0}(x)\,\psi (x)+Q_{2}\overline{\phi }(x)\,\gamma ^{\mu }A_{\mu
}^{0}(x)\,\phi (x),
\end{eqnarray}
and where we have suppressed the kinetic-energy term of the free photon
field. We quantize the theory by imposing equal-time anticommutation rules
for the fermion fields, namely 
\begin{equation}
\left\{ \psi _{\alpha }(\mathbf{x},t),\psi _{\beta }^{\dagger }(\mathbf{y}%
,t)\right\} =\left\{ \phi _{\alpha }(\mathbf{x},t),\phi _{\beta }^{\dagger }(%
\mathbf{y},t)\right\} =\delta _{\alpha \beta }\delta ^{3}\left( \mathbf{x}-%
\mathbf{y}\right) ,
\end{equation}
and all others vanish. In addition, there are the usual commutation rules
for the $A_{0}^{\mu }$ field, and the commutation of the $A_{0}^{\mu }$
field operators with the $\psi $ and $\phi $ field operators.

To specify our notation, we quote the Fourier decomposition of the field
operators, namely 
\begin{equation}
\psi (x)=\sum_{s}\int \frac{d^{3}\mathbf{p}}{\left( 2\pi \right) ^{3/2}}%
\left( \frac{m_{1}}{\omega _{p}}\right) ^{1/2}\left[ b_{\mathbf{p}s}u\left( 
\mathbf{p},s\right) e^{-ip_{1}\cdot x}+d_{\mathbf{p}s}^{\dagger }v\left( 
\mathbf{p},s\right) e^{ip_{1}\cdot x}\right] ,
\end{equation}
with $p_{1}=p_{1}^{\mu }=\left( \omega _{p},\mathbf{p}\right) $, $\omega
_{p}=\sqrt{m_{1}^{2}+\mathbf{p}^{2}}$ and 
\begin{equation}
\phi (x)=\sum_{s}\int \frac{d^{3}\mathbf{p}}{\left( 2\pi \right) ^{3/2}}%
\left( \frac{m_{2}}{\Omega _{p}}\right) ^{1/2}\left[ B_{\mathbf{p}s}U\left( 
\mathbf{p},s\right) e^{-ip_{2}\cdot x}+D_{\mathbf{p}s}^{\dagger }V\left( 
\mathbf{p},s\right) e^{ip_{2}\cdot x}\right] ,
\end{equation}
with $p_{2}=p_{2}^{\mu }=\left( \Omega _{p},\mathbf{p}\right) $, $\Omega
_{p}=\sqrt{m_{2}^{2}+\mathbf{p}^{2}}$. Note that the mass-$m_{1}$
free-particle Dirac spinors $u\left( \mathbf{p},s\right) $, $v\left( \mathbf{%
p},s\right) $,\ which satisfy $\left( \gamma ^{\mu }\widehat{p}_{1\mu
}-m_{1}\right) u\left( \mathbf{p},s\right) =0$ and\ $\left( \gamma ^{\mu }%
\widehat{p}_{1\mu }+m_{1}\right) v\left( \mathbf{p},s\right) =0$, are
normalized such that 
\begin{eqnarray}
u^{\dagger }\left( \mathbf{p},s\right) u\left( \mathbf{p},\sigma \right)
&=&v^{\dagger }\left( \mathbf{p},s\right) v\left( \mathbf{p},\sigma \right) =%
\frac{\omega _{p}}{m_{1}}\delta _{s\sigma }, \\
u^{\dagger }\left( \mathbf{p},s\right) v\left( \mathbf{p},\sigma \right)
&=&v^{\dagger }\left( \mathbf{p},s\right) u\left( \mathbf{p},\sigma \right)
=0.  \notag
\end{eqnarray}
Analogous properties apply to the mass-$m_{2}$ spinors $U$, $V$. The
creation and annihilation operators $b^{\dagger }$, $b$ of the (free)
particles of mass $m_{1}$, and $d^{\dagger }$, $d$ for the corresponding
antiparticles, satisfy the usual anticommutation relations. The
non-vanishing ones are 
\begin{equation}
\left\{ b_{\mathbf{p}s},b_{\mathbf{q}\sigma }^{\dagger }\right\} =\left\{ d_{%
\mathbf{p}s},d_{\mathbf{q}\sigma }^{\dagger }\right\} =\delta _{s\sigma
}\delta ^{3}\left( \mathbf{p}-\mathbf{q}\right) .
\end{equation}
Again, the analogous properties apply to the mass-$m_{2}$ operators $%
B^{\dagger }$, $B$, $D^{\dagger }$, $D$. As a concrete example, we can think
of the mass-$m_{1}$ particles as electrons, and the mass-$m_{2}$ particles
as muons, though any pairs of charged fermions could be considered.

\vskip0.8truecm {\normalsize \noindent {\textbf{\large 3. Variational
principle and adiabatic hypothesis}}}

{\normalsize \vskip0.4truecm }

The Hamiltonian formalism of QFT is based on the covariant eigenvalue
equation 
\begin{equation}
\widehat{P}^{\mu }\,\left| \psi \right\rangle =P^{\mu }\,\left| \psi
\right\rangle ,
\end{equation}
where $\widehat{P}^{\mu }=\left( \widehat{H},\widehat{\mathbf{P}}\right) $
is the energy-momentum operator, $P^{\mu }=\left( E,\mathbf{P}\right) $ is
the corresponding energy-momentum eigenvalue, with $E^{2}-\mathbf{P}^{2}%
\mathbf{=}M^{2}$, where $M$ is the invariant mass of the system. The $\mu =0$%
\ component of (22) cannot be solved for realistic theories, and
approximation methods must be used. We shall develop an approximation method
in this section, based on the variational principle 
\begin{equation}
\delta \left( \left\langle \psi _{h}\right| \widehat{H}_{h}\left( t\right)
-E\left| \psi _{h}\right\rangle _{t=t_{0}}\right) =0,
\end{equation}
which is equivalent to the $\mu =0$\ component of (22), so to this end, we
shall consider the expectation value. The subscript $h$\ is used to stress\
that the equation is written in the Heisenberg representation, which relates
to the Schr\"{o}dinger picture as $\left| \psi _{h}\right\rangle =e^{i%
\widehat{H}_{s}\left( t-t_{0}\right) }\left| \psi _{s}\right\rangle $.

It is well known that variational solutions are only as good as the trial
states that are used. Thus, it is important that the trial states possess as
many features of the exact solution as possible. For our purpose it is
convenient to rewrite equation (23) in the interaction representation 
\begin{equation}
\delta \left( \left\langle \psi _{i}\left( t\right) \right| \widehat{H}%
_{i}\left( t\right) -E\left| \psi _{i}\left( t\right) \right\rangle
_{t=t_{0}}\right) =0,
\end{equation}
where the operator $\widehat{H}_{i}\left( t\right) =\widehat{H}_{0i}\left(
t\right) +\widehat{H}_{Ii}\left( t\right) $ consists on two parts, which
represent the free-field and interacting-field Hamiltonians respectively.
Note that $\widehat{H}_{0i}\left( t\right) =\widehat{H}_{0i}$ is
time-independent.

Using the time-evolution operator $\widehat{U}$ we can express states in the
form 
\begin{equation}
\left| \psi _{i}(t)\right\rangle =\widehat{U}(t,t_{1})\left| \psi
_{i}(t_{1})\right\rangle ,\;\;\;\;\;\;\;\;\;\left\langle \psi _{i}(t)\right|
=\left\langle \psi _{i}(t_{2})\right| \widehat{U}^{\dagger }(t,t_{2}),
\end{equation}
The time-evolution operator $\widehat{U}$ satisfies the familiar
differential equation, 
\begin{equation}
\widehat{H}_{Ii}(t)\widehat{U}(t,t_{0})=i\frac{\partial }{\partial t}%
\widehat{U}(t,t_{0}),
\end{equation}
with the boundary conditions $\widehat{U}(t_{0},t_{0})=I$, where $I$ is the
unit operator. Equation (26) has the iterative solution 
\begin{equation}
\widehat{U}\left( t,t_{0}\right) =\widehat{U}^{\left( 1\right) }\left(
t,t_{0}\right) +\widehat{U}^{\left( 2\right) }\left( t,t_{0}\right) +%
\widehat{U}^{\left( 3\right) }\left( t,t_{0}\right) +...,
\end{equation}
where 
\begin{equation}
\widehat{U}^{\left( 1\right) }\left( t,t_{0}\right) =I,
\end{equation}
\begin{equation}
\widehat{U}^{\left( 2\right) }\left( t,t_{0}\right)
=-i\int_{t_{0}}^{t}dt_{1}\int d^{3}\mathbf{x}_{1}\widehat{\mathcal{H}}%
_{Ii}\left( x_{1}\right) ,
\end{equation}
\begin{equation}
\widehat{U}^{\left( 3\right) }\left( t,t_{0}\right) =\frac{\left( -i\right)
^{2}}{2}\int_{t_{0}}^{t}dt_{1}\int_{t_{0}}^{t}dt_{2}\int d^{3}\mathbf{x}%
_{1}d^{3}\mathbf{x}_{2}T\left( \widehat{\mathcal{H}}_{Ii}\left( x_{1}\right)
\,\widehat{\mathcal{H}}_{Ii}\left( x_{2}\right) \right) ,
\end{equation}
and $\widehat{\mathcal{H}}_{I}$ is defined by (13), (15) and (16).

Substitution of (25) into (24) yields 
\begin{equation}
\delta \left( \left\langle \psi _{i}(t_{2})\right| \widehat{U}^{\dagger
}(t,t_{2})\left( \widehat{H}_{i}\left( t\right) -E\right) \widehat{U}%
(t,t_{1})\left| \psi _{i}(t_{1})\right\rangle _{t=t_{0}}\right) =0.
\end{equation}
Using the properties of the $\widehat{U}$\ operator, $\widehat{U}^{\dagger
}(t,t_{2})=\widehat{U}(t_{2},t)$, and $\widehat{U_{I}}(t_{2},t)\widehat{U_{I}%
}(t,t_{1})=\widehat{U_{I}}(t_{2},t_{1})$ we obtain for $t_{2}>t_{1}$%
\begin{equation}
\delta \left( \left\langle \psi _{i}(t_{2})\right| T\left( \widehat{U}%
(t_{2},t_{1})\left( \widehat{H}_{0i}-E\right) \right) +T\left( \widehat{U}%
(t_{2},t_{1})\widehat{H}_{Ii}\left( t\right) \right) \left| \psi
_{i}(t_{1})\right\rangle _{t=t_{0}}\right) =0,
\end{equation}
where $T$\ is the time-ordering operator. The first term in (32) is 
\begin{equation}
\left\langle \psi _{i}(t_{2})\right| T\left( \widehat{U}(t_{2},t_{1})\left( 
\widehat{H}_{0i}-E\right) \right) \left| \psi _{i}(t_{1})\right\rangle
_{t=t_{0}}=\left\langle \psi _{i}(t_{1})\right| \widehat{H}_{0i}-E\left|
\psi _{i}(t_{1})\right\rangle ,
\end{equation}
since $\widehat{H}_{0i}$ is a time-independent operator, and $\left\langle
\psi _{i}(t_{2})\right| \widehat{U}(t_{2},t_{1})=\left\langle \psi
_{i}(t_{1})\right| $. The second term in (32) can be rearranged as follows, 
\begin{equation}
\left\langle \psi _{i}(t_{2})\right| T\left( \widehat{U}(t_{2},t_{1})%
\widehat{H}_{Ii}\left( t\right) \right) \left| \psi _{i}(t_{1})\right\rangle
_{t=t_{0}}=\left\langle \psi _{i}(t_{1})\right| \widehat{U}%
^{-1}(t_{2},t_{1})T\left( \widehat{U}(t_{2},t_{1})\widehat{H}_{Ii}\left(
t\right) \right) \left| \psi _{i}(t_{1})\right\rangle _{t=t_{0}}
\end{equation}
If $\left| \psi _{i}(t_{2})\right\rangle $ and $\left| \psi
_{i}(t_{1})\right\rangle $ are stationary states, they must coincide to
within an arbitrary phase factor $e^{i\theta }$, which does not affect any
physical values. Hence 
\begin{equation}
\widehat{U}(t_{2},t_{1})\left| \psi _{i}(t_{1})\right\rangle =\left| \psi
_{i}(t_{2})\right\rangle =e^{i\theta }\left| \psi _{i}(t_{1})\right\rangle ,
\end{equation}
where 
\begin{equation}
e^{i\theta }=\left\langle \psi _{i}(t_{1})\right| \widehat{U}%
(t_{2},t_{1})\left| \psi _{i}(t_{1})\right\rangle .
\end{equation}
Thus 
\begin{equation}
\left\langle \psi _{i}(t_{1})\right| \widehat{U}^{-1}(t_{2},t_{1})=e^{-i%
\theta }\left\langle \psi _{i}(t_{1})\right| =\frac{\left\langle \psi
_{i}(t_{1})\right| }{\left\langle \psi _{i}(t_{1})\right| \widehat{U}%
(t_{2},t_{1})\left| \psi _{i}(t_{1})\right\rangle },
\end{equation}
and we obtain 
\begin{equation}
\left\langle \psi _{i}(t_{2})\right| T\left( \widehat{U}(t_{2},t_{1})%
\widehat{H}_{Ii}\left( t\right) \right) \left| \psi _{i}(t_{1})\right\rangle
_{t=t_{0}}=\frac{\left\langle \psi _{i}(t_{1})\right| T\left( \widehat{U}%
(t_{2},t_{1})\widehat{H}_{Ii}\left( t\right) \right) \left| \psi
_{i}(t_{1})\right\rangle _{t=t_{0}}}{\left\langle \psi _{i}(t_{1})\right| 
\widehat{U}(t_{2},t_{1})\left| \psi _{i}(t_{1})\right\rangle }.
\end{equation}
It is known [12], [13] that the numerator of (38) can be written as a
product of two factors, which correspond to all connected and disconnected\
diagrams. The factor, which represents the disconnected diagrams, is
identical with the denominator $\left\langle \psi _{i}(t_{1})\right| 
\widehat{U}(t_{2},t_{1})\left| \psi _{i}(t_{1})\right\rangle $, and
therefore equation (32) takes the final form, 
\begin{equation}
\delta \left( \left\langle \psi _{i}(t_{1})\right| \widehat{H}_{0i}-E\left|
\psi _{i}(t_{1})\right\rangle +\left\langle \psi _{i}(t_{1})\right| T\left( 
\widehat{U}(t_{2},t_{1})\widehat{H}_{Ii}\left( t\right) \right) \left| \psi
_{i}(t_{1})\right\rangle _{t=t_{0}}\right) =0,
\end{equation}
which involves connected diagrams only. This equation is just another form
of (24) and does not contain anything new with respect to it. However,
equation (39) is convenient for the implementation of adiabatic
switching\thinspace \lbrack 12], [13], in which case we put $t_{1}=-\infty $%
. We assume that the Hamiltonian $\widehat{H}_{Ii}\left( t\right) $\ depends
on an interaction constant $\alpha $, hence\ we can switch this interaction
on by the following anzatz 
\begin{equation}
\alpha (t)=\alpha \left( \left( 1-\eta \right) e^{-\varepsilon \left|
t-t_{0}\right| }+\eta \right) ,
\end{equation}
where $\varepsilon $ and $\eta $ are small positive quantities. At time $%
t=-\infty $\ the coupling constant $\alpha \left( -\infty \right) =\alpha
\eta $ is very small (``bait''\ interaction), and becomes the full
interaction at $t=t_{0}$. This means that at $t=-\infty $ the state $\left|
\psi _{i}(-\infty )\right\rangle $ is an eigenstate of a weakly interacting
system, and by process of adiabatic switching, it becomes a stationary state 
$\left| \psi _{i}(t_{0})\right\rangle =\widehat{U}(t_{0},-\infty )\left|
\psi _{i}(-\infty )\right\rangle $ of the fully interacting system. This is
the content of the adiabatic hypothesis. It is evident that the state $%
\left| \psi _{i}(-\infty )\right\rangle $, in a sense,\ contains only
minimal information about the interaction in the system, however it should
include the main features of the ``bait''\ interaction such as kinematical
properties and symmetry of the system. The adiabatic hypothesis permits the
use of simple trial states that take into account the key physical
properties of the system.

\vskip0.8truecm {\normalsize \noindent {\textbf{\large 4. Trial state and
two-fermion system}}}

{\normalsize \vskip0.4truecm }

The simplest Fock-space trial state, that can be written down in the rest
frame ($\mathbf{P}=0$, $E=M$), for a two-fermion system is 
\begin{equation}
\left| \psi _{trial}\right\rangle =\underset{s_{1}s_{2}}{\sum }\int d^{3}%
\mathbf{p}F_{s_{1}s_{2}}(\mathbf{p})b_{\mathbf{p}s_{1}}^{\dagger }D_{-%
\mathbf{p}s_{2}}^{\dagger }\left| 0\right\rangle ,
\end{equation}
where $b_{\mathbf{p}s_{1}}^{\dagger }$ and $D_{-\mathbf{p}s_{2}}^{\dagger }$%
\ are creation operators for fermions of masses $m_{1}$ and $m_{2}$
respectively, and $\left| 0\right\rangle $ is the trial vacuum state such
that $b_{\mathbf{p}s_{1}}\left| 0\right\rangle =D_{\mathbf{p}s_{2}}\left|
0\right\rangle =0$. The $F_{s_{1}s_{2}}$ are four adjustable functions, that
are chosen so that (41) is, as we show in section 5, an eigenstate of the
relativistic total angular momentum operator, its projection, and parity. We
shall use this trial state to implement equation (39), which we write in the
form ($t_{1}=-\infty $, $t_{2}=+\infty $, and $t_{0}=0$) 
\begin{equation}
\delta \left( \left\langle \psi _{trial}\right| \widehat{H}_{0i}-E\left|
\psi _{trial}\right\rangle +\left\langle \psi _{trial}\right| T\left( 
\widehat{U}(+\infty ,-\infty )\widehat{H}_{Ii}\left( t\right) \right) \left|
\psi _{trial}\right\rangle _{t=0}\right) =0.
\end{equation}
In practice $\widehat{U}(+\infty ,-\infty )$\ is represented by the series
(27), however any finite number of terms of the series may be included. In
lowest order ($\widehat{U}=I$) equation (42) reduces to 
\begin{equation}
\delta \left( \left\langle \psi _{trial}\right| \widehat{H}_{0i}-E+\widehat{H%
}_{Ii}\left( t\right) \left| \psi _{trial}\right\rangle _{t=0}\right) =0.
\end{equation}
This equation was used in earlier works [1], [2] (with the trial state (41)
and the interacting Hamiltonian $\widehat{H}_{Ii}=\widehat{H}_{I_{1}i}$
(15)) to obtain bound-state energy spectra to order $\alpha ^{4}$ for all
states of positronium- and muonium-like systems. In this work we shall go
beyond this order.

Note that only the term $\widehat{H}_{I_{1}}$, eq. (15), of the interaction
part of the Hamiltonian contributes in eq. (43), since $\left\langle \psi
_{trial}\right| \widehat{H}_{I_{2}i}\left( t\right) \left| \psi
_{trial}\right\rangle =0$ with the choice (41) of trial state. That is, the
simple trial state (41) does not\ sample the $\widehat{H}_{I_{2}i}$\
interaction term,\ which means that processes that involve the emission or
absorbtion of radiation (i.e., physical photons) are not accomodated. To do
so would require a modification of the trial atate (41). However, this will
not be considered in the present paper, which thereby deals with equations
for pure bound states (or elastic scattering).

The matrix elements (42) to implement the variational principle needed are 
\begin{equation}
\left\langle \psi _{trial}\right| :\widehat{H}_{0}-E:\left| \psi
_{trial}\right\rangle =\underset{s_{1}s_{2}}{\sum }\int d^{3}\mathbf{p}%
F_{s_{1}s_{2}}^{\ast }(\mathbf{p})F_{s_{1}s_{2}}(\mathbf{p})\left( \omega
_{p}+\Omega _{p}-M\right)
\end{equation}
and 
\begin{eqnarray}
&&\left\langle \psi _{trial}\right| :T\left( \widehat{U}(+\infty ,-\infty )%
\widehat{H}_{Ii}\left( t\right) \right) :\left| \psi _{trial}\right\rangle
_{t=0} \\
&=&\left\langle \psi _{trial}\right| :\int d^{4}xT\left( \left( \widehat{U}%
^{\left( 1\right) }(+\infty ,-\infty )+\widehat{U}^{\left( 2\right)
}(+\infty ,-\infty )+...\right) \widehat{\mathcal{H}}_{Ii}\left( x\right)
\delta \left( t\right) \right) :\left| \psi _{trial}\right\rangle  \notag \\
&=&-\frac{q_{1}q_{2}m_{1}m_{2}}{\left( 2\pi \right) ^{3}}\underset{\sigma
_{1}\sigma _{2}s_{1}s_{2}}{\sum }\int \frac{d^{3}\mathbf{p}d^{3}\mathbf{q}}{%
\sqrt{\omega _{p}\omega _{q}\Omega _{p}\Omega _{q}}}F_{s_{1}s_{2}}^{\ast }(%
\mathbf{p})F_{\sigma _{1}\sigma _{2}}(\mathbf{q})\left( -i\right) \widetilde{%
\mathcal{M}}_{s_{1}s_{2}\sigma _{1}\sigma _{2}}\left( \mathbf{p},\mathbf{q}%
\right) ,  \notag
\end{eqnarray}
where $\widetilde{\mathcal{M}}_{s_{1}s_{2}\sigma _{1}\sigma _{2}}\left( 
\mathbf{p},\mathbf{q}\right) $\ is the generalized invariant $\mathcal{M}$\
-matrix 
\begin{equation}
\widetilde{\mathcal{M}}_{s_{1}s_{2}\sigma _{1}\sigma _{2}}\left( \mathbf{p},%
\mathbf{q}\right) =\mathcal{M}_{s_{1}s_{2}\sigma _{1}\sigma _{2}}^{\left(
1\right) }\left( \mathbf{p},\mathbf{q}\right) +\mathcal{M}_{s_{1}s_{2}\sigma
_{1}\sigma _{2}}^{\left( 2\right) }\left( \mathbf{p},\mathbf{q}\right) +..,
\end{equation}
which includes effects in all orders of the interaction, and where the sum
contains all Feynman diagrams, reducible and irreducible. Some details of
the calculations of (45) to order $\mathcal{M}^{\left( 2\right) }$ are
presented in Appendix A.

The variational principle (42) leads to the following equation 
\begin{eqnarray}
&&\sum_{s_{1}s_{2}}\int d^{3}\mathbf{p}\left( \omega _{p}+\Omega
_{p}-E\right) F_{s_{1}s_{2}}(\mathbf{p})\delta F_{s_{1}s_{2}}^{\ast }(%
\mathbf{p}) \\
&&-\frac{m_{1}m_{2}}{\left( 2\pi \right) ^{3}}\underset{\sigma _{1}\sigma
_{2}s_{1}s_{2}}{\sum }\int \frac{d^{3}\mathbf{p}d^{3}\mathbf{q}}{\sqrt{%
\omega _{p}\omega _{q}\Omega _{p}\Omega _{q}}}F_{\sigma _{1}\sigma _{2}}(%
\mathbf{q})\left( -i\right) \widetilde{\mathcal{M}}_{s_{1}s_{2}\sigma
_{1}\sigma _{2}}\left( \mathbf{p,q}\right) \delta F_{s_{1}s_{2}}^{\ast }(%
\mathbf{p})=0.  \notag
\end{eqnarray}

We now discuss the structure of the $\widetilde{\mathcal{M}}$- matrix on the
basis of the\ expansion (27) for $\widehat{U}$, and restrict our
consideration to one-loop level only. In lowest order, $\widehat{U}=\widehat{%
U}^{\left( 1\right) }=I$, we obtain 
\begin{eqnarray}
\mathcal{M}_{s_{1}s_{2}\sigma _{1}\sigma _{2}}^{\left( 1\right) }\left( 
\mathbf{p},\mathbf{q}\right) &\equiv &\mathcal{M}_{s_{1}s_{2}\sigma
_{1}\sigma _{2}}^{ope}\left( \mathbf{p,q}\right) \\
&=&-\overline{u}\left( \mathbf{p},s_{1}\right) \left( -iQ_{1}\gamma ^{\mu
}\right) u\left( \mathbf{q},\sigma _{1}\right) iD_{\mu \nu }(p-q)\overline{V}%
\left( -\mathbf{q},\sigma _{2}\right) \left( -iQ_{2}\gamma ^{\nu }\right)
V\left( -\mathbf{p},s_{2}\right) ,  \notag
\end{eqnarray}
where $\mathcal{M}_{s_{1}s_{2}\sigma _{1}\sigma _{2}}^{ope}\left( \mathbf{p,q%
}\right) $ is the usual invariant matrix element corresponding to the
one-photon exchange Feynman diagram obtained and considered in [2]. This
expression involves the Fourier transform of the Green's function (eq. (7)),
namely 
\begin{equation}
D_{\mu \nu }(x-x^{\prime })=\int \frac{d^{4}k}{(2\pi )^{4}}D_{\mu \nu
}(k)e^{-ik(x-x^{\prime })}.
\end{equation}
The Green's function $D_{\mu \nu }(p-q)$\ consists of two parts 
\begin{equation}
D_{\mu \nu }(p-q)=\frac{1}{2}\left( D_{\mu \nu }(p_{1}-q_{1})+D_{\mu \nu
}(q_{2}-p_{2})\right) ,
\end{equation}
where $D_{\mu \nu }(p_{1}-q_{1})$ and $D_{\mu \nu }(q_{2}-p_{2})$ are the
Green's functions of particles with masses $m_{1}$\ and $m_{2}$\
respectively. Note that the four-vectors $p_{1}$, $p_{2}$, $q_{1}$\ and $%
q_{2}$\ are defined as in formulae (18), (19). For a fermion-antifermion
system like positronium we obtain [1] the additional virtual-annihilation
term ($Q_{1}=Q_{2}\equiv e$) for $\mathcal{M}_{s_{1}s_{2}\sigma _{1}\sigma
_{2}}^{\left( 1\right) }$: 
\begin{equation}
\mathcal{M}_{s_{1}s_{2}\sigma _{1}\sigma _{2}}^{ann}\left( \mathbf{p,q}%
\right) =\overline{u}\left( \mathbf{p},s_{1}\right) \left( -ie\gamma ^{\mu
}\right) v\left( -\mathbf{p},s_{2}\right) iD_{\mu \nu }\left( \omega
_{p}\right) \overline{v}\left( -\mathbf{q},\sigma _{2}\right) \left(
-ie\gamma ^{\nu }\right) u\left( \mathbf{q},\sigma _{1}\right) .
\end{equation}
The terms of next order in the coupling require the inclusion of the
operator $\widehat{U}^{\left( 2\right) }$, eq. (29), and can be written in
the form 
\begin{equation}
\mathcal{M}_{s_{1}s_{2}\sigma _{1}\sigma _{2}}^{\left( 2\right) }\equiv
\sum_{i=1,2}\left( \mathcal{M}_{s_{1}s_{2}\sigma _{1}\sigma _{2}}^{vac_{i}}+%
\mathcal{M}_{s_{1}s_{2}\sigma _{1}\sigma _{2}}^{ver_{i}}+\mathcal{M}%
_{s_{1}s_{2}\sigma _{1}\sigma _{2}}^{mass_{i}^{\prime }}+\mathcal{M}%
_{s_{1}s_{2}\sigma _{1}\sigma _{2}}^{mass_{i}^{\prime \prime }}\right) +%
\mathcal{M}_{s_{1}s_{2}\sigma _{1}\sigma _{2}}^{2\gamma },
\end{equation}
where the index $i=1,2$ corresponds to masses $m_{1}$\ and $m_{2}$
respectively. One-loop level radiative corrections include second-order
vacuum polarization $\mathcal{M}^{vac}$, vertex corrections $\mathcal{M}%
^{ver}$, and the fermion self-energy $\mathcal{M}^{mass}$. Their explicit
form is given by 
\begin{eqnarray}
&&\mathcal{M}_{s_{1}s_{2}\sigma _{1}\sigma _{2}}^{\left( 2\right)
vac_{1}}\left( \mathbf{p,q}\right) \\
&=&-\overline{u}\left( \mathbf{p},s_{1}\right) \left( -iQ_{1}\gamma ^{\mu
}\right) u\left( \mathbf{q},\sigma _{1}\right) iD_{\mu \nu }\left(
p-q\right) \mathbf{\Pi }^{\nu \alpha }\left( p_{1}-q_{1}\right) \times 
\notag \\
&&\times iD_{\alpha \beta }\left( p-q\right) \overline{V}\left( -\mathbf{q}%
,\sigma _{2}\right) \left( -iQ_{2}\gamma ^{\beta }\right) V\left( -\mathbf{p}%
,s_{2}\right) ,  \notag \\
&&  \notag \\
&&\mathcal{M}_{s_{1}s_{2}\sigma _{1}\sigma _{2}}^{\left( 2\right)
ver_{1}}\left( \mathbf{p,q}\right) \\
&=&-\overline{u}\left( \mathbf{p},s_{1}\right) \left( -iQ_{1}\right) \mathbf{%
\Lambda }^{\alpha }\left( p_{1},q_{1}\right) u\left( \mathbf{q},\sigma
_{1}\right) iD_{\alpha \beta }\left( p-q\right) \overline{V}\left( -\mathbf{q%
},\sigma _{2}\right) \left( -iQ_{2}\gamma ^{\beta }\right) V\left( -\mathbf{p%
},s_{2}\right) ,  \notag \\
&&  \notag \\
&&\mathcal{M}_{s_{1}s_{2}\sigma _{1}\sigma _{2}}^{\left( 2\right)
mass_{1}^{\prime }}\left( \mathbf{p,q}\right) \\
&=&-\overline{u}\left( \mathbf{p},s_{1}\right) \mathbf{\Sigma }\left(
p_{1}\right) S_{\psi }\left( p_{1}\right) \left( -iQ_{1}\gamma ^{\alpha
}\right) u\left( \mathbf{q},\sigma _{1}\right) iD_{\alpha \beta }\left(
p-q\right) \overline{V}\left( -\mathbf{q},\sigma _{2}\right) \left(
-iQ_{2}\gamma ^{\beta }\right) V\left( -\mathbf{p},s_{2}\right) ,  \notag \\
&&  \notag \\
&&\mathcal{M}_{s_{1}s_{2}\sigma _{1}\sigma _{2}}^{\left( 2\right)
mass_{1}^{\prime \prime }}\left( \mathbf{p,q}\right) \\
&=&-\overline{u}\left( \mathbf{p},s_{1}\right) \left( -iQ_{1}\gamma ^{\alpha
}\right) S_{\psi }\left( p_{1}\right) \mathbf{\Sigma }\left( p_{1}\right)
u\left( \mathbf{q},\sigma _{1}\right) iD_{\alpha \beta }\left( p-q\right) 
\overline{V}\left( -\mathbf{q},\sigma _{2}\right) \left( -iQ_{2}\gamma
^{\beta }\right) V\left( -\mathbf{p},s_{2}\right) ,  \notag
\end{eqnarray}
where the standard definitions of the $\mathbf{\Pi }$, $\mathbf{\Lambda }$
and\ $\mathbf{\Sigma }$ functions apply (we display functions and operators
for the $\psi $ field only), namely 
\begin{equation}
\mathbf{\Pi }^{\nu \alpha }\left( p_{1}-q_{1}\right) =-iQ_{1}^{2}\int \frac{%
d^{4}k}{\left( 2\pi \right) ^{4}}Tr\left[ \gamma ^{\nu }S_{\psi }\left(
k+p_{1}-q_{1}\right) \gamma ^{\alpha }S_{\psi }\left( k\right) \right] ,
\end{equation}
for the vacuum polarization function, 
\begin{equation}
\mathbf{\Lambda }^{\alpha }\left( p_{1},q_{1}\right) =-iQ_{1}^{2}\int \frac{%
d^{4}k}{\left( 2\pi \right) ^{4}}\gamma ^{\nu }S_{\psi }\left(
k+p_{1}\right) \gamma ^{\alpha }S_{\psi }\left( k+q_{1}\right) \gamma ^{\mu
}D_{\mu \nu }\left( k\right) ,
\end{equation}
for the vertex function, and 
\begin{equation}
\mathbf{\Sigma }\left( p_{1}\right) =Q_{1}^{2}\int \frac{d^{4}k}{\left( 2\pi
\right) ^{4}}D_{\mu \nu }\left( k\right) \gamma ^{\mu }S_{\psi }\left(
p_{1}-k\right) \gamma ^{\nu },
\end{equation}
for the self-energy function. The fermion propagator is the usual form 
\begin{equation}
iS_{\psi }\left( x_{1}^{\prime }-x^{\prime }\right) =\left\langle 0\right|
T\psi (x_{1}^{\prime })\overline{\psi }(x^{\prime })\left| 0\right\rangle .
\end{equation}
Its Fourier transform is defined by 
\begin{equation}
iS_{\psi }\left( x_{1}^{\prime }-x^{\prime }\right) =\int \frac{d^{4}k}{%
\left( 2\pi \right) ^{4}}e^{-ik\left( x_{1}^{\prime }-x^{\prime }\right)
}S_{\psi }\left( k\right) =\int \frac{d^{4}k}{\left( 2\pi \right) ^{4}}%
e^{-ik\left( x_{1}^{\prime }-x^{\prime }\right) }\frac{i\left( \gamma ^{\mu
}k_{\mu }+mI\right) }{k^{2}-m^{2}+i0}.
\end{equation}
The two-photon exchange matrix element includes box and crossed-box matrix
element 
\begin{equation}
\mathcal{M}_{s_{1}s_{2}\sigma _{1}\sigma _{2}}^{2\gamma }\left( \mathbf{p,q}%
\right) =\mathcal{M}_{s_{1}s_{2}\sigma _{1}\sigma _{2}}^{box}\left( \mathbf{%
p,q}\right) +\mathcal{M}_{s_{1}s_{2}\sigma _{1}\sigma _{2}}^{c-box}\left( 
\mathbf{p,q}\right) ,
\end{equation}
where 
\begin{eqnarray}
&&\mathcal{M}_{s_{1}s_{2}\sigma _{1}\sigma _{2}}^{box}\left( \mathbf{p,q}%
\right) \\
&=&-\int \frac{d^{4}k}{2\pi }\,\overline{u}\left( \mathbf{p},s_{1}\right)
\left( -iQ_{1}\Gamma _{\psi }^{\mu \alpha }\left( p_{1}-k\right) \right)
u\left( \mathbf{q},\sigma _{1}\right) iD_{\mu \nu \alpha \beta }\overline{V}%
\left( -\mathbf{q},\sigma _{2}\right) \left( -iQ_{2}\Gamma _{\phi }^{\beta
\nu }\left( p_{2}-k\right) \right) V\left( -\mathbf{p},s_{2}\right) ,  \notag
\end{eqnarray}
\begin{eqnarray}
&&\mathcal{M}_{s_{1}s_{2}\sigma _{1}\sigma _{2}}^{c-box}\left( \mathbf{p,q}%
\right) \\
&=&-\int \frac{d^{4}k}{2\pi }\,\overline{u}\left( \mathbf{p},s_{1}\right)
\left( -iQ_{1}\Gamma _{\psi }^{\mu \alpha }\left( p_{1}-k\right) \right)
u\left( \mathbf{q},\sigma _{1}\right) iD_{\mu \nu \alpha \beta }\overline{V}%
\left( -\mathbf{q},\sigma _{2}\right) \left( -iQ_{2}\Gamma _{\phi }^{\nu
\beta }\left( q_{2}+k\right) \right) V\left( -\mathbf{p},s_{2}\right) . 
\notag
\end{eqnarray}
In the above we used the notation 
\begin{eqnarray}
D_{\mu \nu \alpha \beta } &=&D_{\mu \nu }\left( k\right) \left( D_{\alpha
\beta }\left( p_{1}-q_{1}-k\right) +D_{\beta \alpha }\left(
p_{2}-q_{2}-k\right) \right) , \\
\Gamma _{\psi }^{\mu \alpha }\left( p_{1}-k\right) &=&\gamma ^{\mu }S_{\psi
}\left( p_{1}-k\right) \gamma ^{\alpha }, \\
\Gamma _{\phi }^{\beta \nu }\left( p_{2}-k\right) &=&\gamma ^{\beta }S_{\phi
}\left( p_{2}-k\right) \gamma ^{\nu }, \\
\Gamma _{\phi }^{\nu \beta }\left( q_{2}+k\right) &=&\gamma ^{\nu }S_{\phi
}\left( q_{2}+k\right) \gamma ^{\beta }.
\end{eqnarray}
The one-loop renormalization scheme in our formalism is considered in
Appendix B.

Note that the $\mathcal{M}$-matrix arises naturally in this formalism, i.e., 
$\mathcal{M}$ is not put in by hand, nor does its derivation require
additional Fock-space terms in the variational trial state (41), as is the
case in previous formulations (e.g. [14], [15]).

It is of interest to show that in the non-relativistic case the variational
equation (47) reduces to the Schr\"{o}dinger equation. Indeed, in the
non-relativistic limit, the functions $F_{s_{1}s_{2}}$ can be written as 
\begin{equation}
F_{s_{1}s_{2}}(\mathbf{p})=F(\mathbf{p})\Lambda _{s_{1}s_{2}},
\end{equation}
where the non-zero elements of $\Lambda _{ij}$ for total spin singlet ($S=0$%
) states are $\Lambda _{12}=-\Lambda _{21}=\frac{1}{\sqrt{2}}$, while for
the spin triplet ($S=1$)\ states the non-zero elements are $\Lambda _{11}=1$
for $m_{s}=+1,$ $\Lambda _{12}=\Lambda _{21}=\frac{1}{\sqrt{2}}$ for $%
m_{s}=0 $, and $\Lambda _{22}=1$ for $m_{s}=-1$. We use the notation that
the subscripts 1 and 2 of $\Lambda $ correspond to $m_{s}=1/2$ and $%
m_{s}=-1/2$ (or $\uparrow $\ and\ $\downarrow $) respectively. Substituting
(69) into (47), multiplying the result by $\Lambda _{s_{1}s_{2}}$ and
summing over $s_{1}$ and $s_{2}$, gives the eigenvalue equation, which
determines the two-particle energy $E$ 
\begin{equation}
(\omega (\mathbf{p})+\Omega (\mathbf{p})-E)F(\mathbf{p})=\frac{1}{(2\pi )^{3}%
}\int d^{3}\mathbf{q\,}\mathcal{K}(\mathbf{p},\mathbf{q})F(\mathbf{q}),
\end{equation}
where 
\begin{equation}
\mathcal{K}(\mathbf{p},\mathbf{q})=-i\sum_{s_{1}s_{2}\sigma _{1}\sigma
_{2}}\Lambda _{s_{1}s_{2}}\mathcal{M}_{s_{1}s_{2}\sigma _{1}\sigma
_{2}}^{\left( 1\right) }\left( \mathbf{p,q}\right) \Lambda _{\sigma
_{1}\sigma _{2}}.
\end{equation}
To lowest-order in $\left( \left| \mathbf{p}\right| \mathbf{,}\left| \mathbf{%
q}\right| \right) /\left( m_{1},m_{2}\right) $ (i.e. in the non-relativistic
limit), the kernel (71) reduces to $\mathcal{K}=q_{1}q_{2}/\left| \mathbf{p-q%
}\right| ^{2}$, and so (70) reduces to the (momentum-space) Schr\"{o}dinger
equation 
\begin{equation}
\left( \frac{\mathbf{p}^{2}}{2m_{r}}-{\LARGE \varepsilon }\right) F(\mathbf{p%
})=\frac{Q_{1}Q_{2}}{(2\pi )^{3}}\int d^{3}\mathbf{q}\frac{1}{\left| \mathbf{%
p-q}\right| ^{2}}F(\mathbf{q}),
\end{equation}
where $\varepsilon =E-\left( m_{1}+m_{2}\right) $ and $m_{r}=m_{1}m_{2}/%
\left( m_{1}+m_{2}\right) $ is the reduced mass. This verifies that the
relativistic two-fermion equation (47) has the required non-relativistic
limit.

\vskip0.8truecm {\normalsize \noindent {\textbf{\large 5. Partial-wave
decomposition and classification of states}}} {\normalsize \vskip0.4truecm }

In the relativistic case we shall not complete the variational procedure in
(47) at this stage to obtain final equations for the four functions $%
F_{s_{1}s_{2}}$, because they are not independent in general. We require
that the trial state must be an eigenstate of the relativistic total angular
momentum operator, its projection, and parity, i.e., 
\begin{equation}
\left[ 
\begin{array}{c}
\widehat{\mathbf{J}}^{2} \\ 
\widehat{J}_{3} \\ 
\widehat{\mathcal{P}}
\end{array}
\right] \,\left| \psi _{trial}\right\rangle =\left[ 
\begin{array}{c}
J\left( J+1\right) \\ 
m_{J} \\ 
P
\end{array}
\right] \,\left| \psi _{trial}\right\rangle .
\end{equation}

The total angular momentum operator is defined by the expression{\normalsize %
\ 
\begin{equation}
\widehat{\mathbf{J}}=\int d^{3}\mathbf{x\,}\psi ^{\dagger }\left( x\right) 
\mathbf{(}\widehat{\mathbf{L}}+\widehat{\mathbf{S}})\psi \left( x\right)
+\int d^{3}\mathbf{x\,}\phi ^{\dagger }\left( x\right) \mathbf{(}\widehat{%
\mathbf{L}}+\widehat{\mathbf{S}})\phi \left( x\right) ,
\end{equation}
}where $\widehat{\mathbf{L}}$ is the orbital angular momentum and $\widehat{%
\mathbf{S}}$ - the spin operator: $\widehat{\mathbf{L}}=\widehat{\mathbf{x}}%
\times \widehat{\mathbf{p}}$ and $\widehat{\mathbf{S}}=\frac{1}{2}\widehat{%
\overrightarrow{\sigma }}$. Using the field operators $\psi \left( x\right) $
and $\phi \left( x\right) $ in the form (18), (19), we obtain\ after some
calculations\ the expression for operator $\widehat{\mathbf{J}}$. Explicit
forms for the operators $\widehat{\mathbf{J}}^{2}$, $\widehat{J}_{3}$ are
given in papers [1], [2].

For a particle-antiparticle\ system charge conjugation invariance represents
an additional requirement, i.e., we wish to construct the eigenvalues of 
\begin{equation}
\widehat{\mathcal{C}}\,\left| e^{+}e^{-}\right\rangle =C\,\left|
e^{+}e^{-}\right\rangle .
\end{equation}
However, this does not apply to the unequal mass case $m_{1}\neq m_{2}$.\
The functions $F_{s_{1}s_{2}}(\mathbf{p})$ can be written in the general
form 
\begin{equation}
F_{s_{1}s_{2}}(\mathbf{p})=\sum_{\ell
_{s_{1}s_{2}}}\sum_{m_{s_{1}s_{2}}}f_{s_{1}s_{2}}^{\ell
_{s_{1}s_{2}}m_{s_{1}s_{2}}}\left( p\right) Y_{\ell
_{s_{1}s_{2}}}^{m_{s_{1}s_{2}}}(\widehat{\mathbf{p}}),
\end{equation}
where $Y_{\ell _{s_{1}s_{2}}}^{m_{s_{1}s_{2}}}(\widehat{\mathbf{p}})$\ are
the usual spherical harmonics. Here and henceforth we will use the notation $%
p=\left| \mathbf{p}\right| $ etc. (four-vectors will be written as $p^{\mu }$%
). The orbital indices $\ell _{s_{1}s_{2}}$and $m_{s_{1}s_{2}}$ and the
radial functions $f_{s_{1}s_{2}}^{\ell _{s_{1}s_{2}}m_{s_{1}s_{2}}}\left(
p\right) $ depend on the spin variables $s_{1}$ and $s_{2}$. Substitution of
(76) into (41) and then into (73) leads to two categories of relations among
the adjustable functions $F_{s_{1}s_{2}}(\mathbf{p})$.

\vskip0.2truecm

{\normalsize \noindent }\textbf{Mixed-spin states }

In this case $\ell _{s_{1}s_{2}}\equiv \ell =J$ and the general solution of
the system (79) is 
\begin{equation}
F_{s_{1}s_{2}}(\mathbf{p})=C_{1}F_{s_{1}s_{2}}^{\left( sg\right) }(\mathbf{p}%
)+C_{2}F_{s_{1}s_{2}}^{\left( tr\right) }(\mathbf{p}),
\end{equation}
where $C_{1}$ and $C_{2}$ are arbitrary constants. $F_{s_{1}s_{2}}^{\left(
sg\right) }(\mathbf{p})$ and $F_{s_{1}s_{2}}^{\left( tr\right) }(\mathbf{p})$
are functions, which correspond to pure singlet states with total spin $S=0$
and triplet states with $S=1$ respectively. The singlet functions have the
form 
\begin{equation}
F_{s_{1}s_{2}}^{\left( sg\right) }(\mathbf{p})=C^{\left( sg\right)
m_{s_{1}s_{2}}}f^{J}(p)Y_{J}^{m_{s_{1}s_{2}}}(\widehat{\mathbf{p}}).
\end{equation}
where the Clebsch-Gordan (C-G) coefficients $C^{\left( sg\right)
m_{s_{1}s_{2}}}$\ are: $C^{\left( sg\right) m_{11}}=C^{\left( sg\right)
m_{22}}=0$,\ $C^{\left( sg\right) m_{12}}=-C^{\left( sg\right) m_{21}}=1$.
For the quantum numbers $m_{s_{1}s_{2}}$ one obtains: $m_{11}=m_{22}=0\ $and$%
\;m_{12}=m_{21}=m_{J}$.\ The spin and radial variables separate in the sense
that the factors $f_{s_{1}s_{2}}^{\left( sg\right) J}(p)$ have a common
radial function $f^{J}(p)$.

The triplet functions have the form 
\begin{equation}
F_{s_{1}s_{2}}^{\left( tr\right) }(\mathbf{p})=C_{Jm_{J}}^{\left( tr\right)
Jm_{s}}f^{J}(p)Y_{J}^{m_{s_{1}s_{2}}}(\widehat{\mathbf{p}}),
\end{equation}
where $C_{Jm_{J}}^{\left( tr\right) Jm_{s}}$ are the C-G coefficients for $%
S=1$, and 
\begin{equation}
m_{11}=m_{J}-1,\;\ \ \ m_{12}=m_{21}=m_{J},\;\ \ \ m_{22}=m_{J}+1.
\end{equation}
We need to note that (78) is true for the singlet states $J\geq 0$, while
(79) is true for the triplet states $J\geq 1$. Thus, the coefficient $C_{2}$
in (77) is zero when $J=0$. In other words, for $J=0$, only the pure singlet
state arises. For a system-like positronium the requirement (75) decouples
the singlet and triplet states for all $J$. Indeed, the charge conjugation
eigenstates are 
\begin{equation}
\left| sg\right\rangle =\sum_{_{s_{1}s_{2}}}C_{Jm_{J}}^{\left( sg\right)
m_{s_{1}s_{2}}}\int d^{3}\mathbf{p\,}f^{J}(p)Y_{J}^{m_{J}}(\widehat{\mathbf{p%
}})b_{\mathbf{p}s_{1}}^{\dagger }d_{-\mathbf{p}s_{2}}^{\dagger }\left|
0\right\rangle
\end{equation}
with $C=\left( -1\right) ^{J}$ for the pure singlet states, and 
\begin{equation}
\left| tr\right\rangle =\sum_{_{s_{1}s_{2}}}C_{Jm_{J}}^{\left( tr\right)
m_{s_{1}s_{2}}}\int d^{3}\mathbf{p\,}f^{J}(p)Y_{J}^{m_{J}}(\widehat{\mathbf{p%
}})b_{\mathbf{p}s_{1}}^{\dagger }d_{-\mathbf{p}s_{2}}^{\dagger }\left|
0\right\rangle
\end{equation}
with $C=\left( -1\right) ^{J+1}$ for the pure triplet states, as it
discussed in Appendix A.

The states (81) and (82) diagonalize the Hamiltonian (13). Thus, for
positronium-like systems, the states can be characterized by the spin
quantum number $S$, and the mixed states (77) separate into singlet states
(parastates $S=0$) and triplet states (orthostates $S=1$). For distinct
particles ($m_{1}\neq m_{2}$) $C$ is not conserved and there is no
separation into pure singlet and triplet states in general. Thus for
arbitrary mass ratio we need to diagonalize the expectation value of the QFT
Hamiltonian $\Delta \widehat{H}=\widehat{H}-\widehat{H}_{NR}-M$, in the
basis of the states $\left| sg\right\rangle $, $\left| tr\right\rangle $\
with (78) and (79) respectively for $J\neq 0$.\ This can be achieved by the
following linear transformation\ 
\begin{equation}
\left[ 
\begin{array}{c}
\left| sg_{q}\right\rangle \\ 
\left| tr_{q}\right\rangle
\end{array}
\right] =U\left[ 
\begin{array}{c}
\left| sg\right\rangle \\ 
\left| tr\right\rangle
\end{array}
\right] ,
\end{equation}
where $U$ is a unimodular matrix $U_{11}=U_{22}\equiv
a,\;U_{12}=-U_{21}\equiv -b$, the components of which are, evidently,
defined by the dynamics of the system. The new states, which diagonalize the
expectation value of $\widehat{H}$, shall be called quasi-singlet $\mid
sg_{q}\rangle $ and quasi-triplet $\mid tr_{q}\rangle $ states 
\begin{eqnarray}
\left| sg_{q}\right\rangle &=&\sum_{_{s_{1}s_{2}}}C_{Jm_{J}}^{\left(
s_{s}\right) Jm_{s_{1}s_{2}}}\int d^{3}\mathbf{p\,}f^{J}(p)Y_{J}^{m_{J}}(%
\widehat{\mathbf{p}})b_{\mathbf{p}s_{1}}^{\dagger }D_{-\mathbf{p}%
s_{2}}^{\dagger }\left| 0\right\rangle ,  \notag \\
&& \\
\left| tr_{q}\right\rangle &=&\sum_{_{s_{1}s_{2}}}C_{Jm_{J}}^{\left(
s_{t}\right) Jm_{s_{1}s_{2}}}\int d^{3}\mathbf{p\,}f^{J}(p)Y_{J}^{m_{J}}(%
\widehat{\mathbf{p}})b_{\mathbf{p}s_{1}}^{\dagger }D_{-\mathbf{p}%
s_{2}}^{\dagger }\left| 0\right\rangle ,  \notag
\end{eqnarray}
where the coefficients $C_{Jm_{J}}^{\left( s_{s}\right) Jm_{s_{1}s_{2}}}$
and $C_{Jm_{J}}^{\left( s_{t}\right) Jm_{s_{1}s_{2}}}$ satisfy the following
condition 
\begin{equation}
\sum_{\nu _{1}\nu _{2}m_{J}}\left( C_{Jm_{J}}^{\left( s_{s}\right) Jm_{\nu
}}\right) ^{2}=\sum_{\nu _{1}\nu _{2}m_{J}}\left( C_{Jm_{J}}^{\left(
s_{t}\right) Jm_{\nu }}\right) ^{2}=2\left( 2J+1\right) .
\end{equation}
Note that these coefficients differ from C-G coefficients, due to the nature
of the coupling. These coupled quasi-states arise only for $J>0$. For $J=0$
pure $S=0$ states occur. Quasi-singlet and quasi-triplet states are both
characterized by the same quantum numbers $J$, $m_{J}$ and $P=(-1)^{J+1}$.
Due to the unimodularity of the matrix $U$ we can identify quasi-singlet and
quasi-triplet states by quasi-spin (like isospin) $t=1/2$ with $t_{3}=\mp
1/2 $, which is a new quantum number. However, for our purpose it is more
convenient to use the value $s=t_{3}+1/2$, which gives $s=s_{s}=0$ or $%
s=s_{t}=1$ for quasi-singlet and quasi-triplet states respectively. In this
case the labels $s_{s}$ and $s_{t}$ reflect better the meaning of the
indicated\ quasi-states. As was shown in [2], for positronium the
quasi-states become true singlet and triplet states with different charge
conjugation quantum numbers. It is useful to note for subsequent
calculations that the coefficients $C_{1}$ and $C_{2}$ in (77) are $C_{1}=a$%
, $C_{2}=-b$ for quasi-singlet states $s=s_{s}=0$, and $C_{1}=b$, $C_{2}=a$
for quasi-triplet states $s=s_{t}=1$.

\vskip0.2truecm

{\normalsize \noindent }\textbf{The triplet }$\ell $-\textbf{mixing states}

These states occur for $\ell _{s_{1}s_{2}}\equiv \ell =J\mp 1$. The
adjustable functions have the form 
\begin{equation}
F_{s_{1}s_{2}}(\mathbf{p})=C_{Jm_{J}}^{\left( tr\right)
(J-1)m_{s}}f^{J-1}(p)Y_{J-1}^{m_{s_{1}s_{2}}}(\widehat{\mathbf{p}}%
)+C_{Jm_{J}}^{\left( tr\right) (J+1)m_{s}}f^{J+1}(p)Y_{J+1}^{m_{s_{1}s_{2}}}(%
\widehat{\mathbf{p}}),
\end{equation}
where the coefficients $C_{Jm_{J}}^{\left( tr\right) (J\mp 1)m_{s}}$ are
precisely the C-G coefficients. Expression (86) involves two radial
functions $f^{J-1}(p)$ and $f^{J+1}(p)$ which correspond to $\ell =J-1$ and $%
\ell =J+1$. This reflects the fact that the orbital angular momentum is not
conserved and $\ell $ is not a good quantum number. The system in these
states is characterized by $J,$ $m_{J},$ and $P=(-1)^{J}$. In spectroscopic
notation $^{2S+1}L_{J}$, these states are a mixture of $^{3}\left(
J-1\right) _{J}$, and $^{3}\left( J+1\right) _{J}$\ states. The exception is
the state with $J=0$, for which the orbital angular momentum is conserved.
Indeed, for $J=0$\ the function $f^{J-1}(p)$ does not exist (see Appendix
A), thus the function $F_{s_{1}s_{2}}(\mathbf{p})$ is defined only by the
second term in (86). Note that $\ell $-mixing states appear for principal
quantum number $n\geq 3$\ only.

For practical applications it is convenient to express pure states through
the Dirac's $\Gamma $\ matrices, namely 
\begin{equation}
F_{s_{1}s_{2}}\left( \mathbf{p}\right) =\sum_{i=1}^{3}f^{\ell }(p)\overline{u%
}_{\mathbf{p}s_{1}}\Gamma _{i}^{\ell }\left( \widehat{\mathbf{p}}\right) V_{-%
\mathbf{p}s_{2}}.
\end{equation}

The form of $\Gamma _{i}^{\ell }\left( \widehat{\mathbf{p}}\right) $\
depends on the particular states and the index $i$\ corresponds to three
cases: $i=1$, when $m_{\ell }=m_{J}-1$, $i=2$, when $m_{\ell }=m_{J}$, and $%
i=3$, when $m_{\ell }=m_{J}+1$. It is given below for the following cases:

\vskip0.1truecm

{\normalsize \noindent }\textbf{Pure Singlet States} $\ell =J,\;\ J\geq
0,\;\;P=(-1)^{J+1}$

\begin{equation}
\Gamma _{i}^{J}\left( \widehat{\mathbf{p}}\right) \equiv \Gamma ^{\ell
}\left( \widehat{\mathbf{p}}\right) =\gamma ^{5}Y_{J}^{m_{J}}\left( \widehat{%
\mathbf{p}}\right) ,
\end{equation}
that is 
\begin{equation}
\left| \psi _{trial}\right\rangle =\sum_{s_{1}s_{2}}\int d^{3}\mathbf{p\,}%
f^{J}(p)Y_{J}^{m_{J}}\left( \widehat{\mathbf{p}}\right) \overline{u}_{%
\mathbf{p}s_{1}}\gamma ^{5}V_{-\mathbf{p}s_{2}}b_{\mathbf{p}s_{1}}^{\dagger
}D_{-\mathbf{ps}_{2}}^{\dagger }\left| 0\right\rangle
\end{equation}
\vskip0.1truecm

{\normalsize \noindent }\textbf{Pure Triplet States} $\ell =J,\;\
J>0,\;\;P=(-1)^{J+1}$

\vskip0.1truecm

When $m_{\ell }=m_{J}-1$%
\begin{eqnarray}
\Gamma _{1}^{J}\left( \widehat{\mathbf{p}}\right)  &=&\frac{1}{2}\left( 
\frac{m_{1}m_{2}}{\left( \omega _{p}+m_{1}\right) \left( \Omega
_{p}+m_{2}\right) }\right) ^{1/2}\left( \frac{\left( J+m_{J}\right)
(J-m_{J}+1)}{J\left( J+1\right) }\right) ^{1/2}  \notag \\
&&\times \left( \gamma ^{1}+i\gamma ^{2}-i\left( \sigma ^{01}+i\sigma
^{02}\right) \right) Y_{J}^{m_{J}-1}\left( \widehat{\mathbf{p}}\right) ,
\end{eqnarray}
when $m_{\ell }=m_{J},$%
\begin{equation}
\Gamma _{2}^{J}\left( \widehat{\mathbf{p}}\right) =-\left( \frac{m_{1}m_{2}}{%
\left( \omega _{p}+m_{1}\right) \left( \Omega _{p}+m_{2}\right) }\right)
^{1/2}\frac{m_{J}}{\left( J\left( J+1\right) \right) ^{1/2}}\left( \gamma
^{3}-i\sigma ^{03}\right) Y_{J}^{m_{J}}\left( \widehat{\mathbf{p}}\right) ,
\end{equation}
and $m_{\ell }=m_{J}+1,$%
\begin{eqnarray}
\Gamma _{3}^{J}\left( \widehat{\mathbf{p}}\right)  &=&\frac{1}{2}\left( 
\frac{m_{1}m_{2}}{\left( \omega _{p}+m_{1}\right) \left( \Omega
_{p}+m_{2}\right) }\right) ^{1/2}\left( \frac{\left( J-m_{J}\right)
(J+m_{J}+1)}{J\left( J+1\right) }\right) ^{1/2}  \notag \\
&&\times \left( \gamma ^{1}-i\gamma ^{2}-i\left( \sigma ^{01}-i\sigma
^{02}\right) \right) Y_{J}^{m_{J}+1}\left( \widehat{\mathbf{p}}\right) .
\end{eqnarray}
\vskip0.1truecm

{\normalsize \noindent }\textbf{Pure Triplet States} $\ell =J-1,\;\ J\geq
1,\;\;P=(-1)^{J}$

\vskip0.1truecm

When $m_{\ell }=m_{J}-1$%
\begin{eqnarray}
\Gamma _{1}^{J-1}\left( \widehat{\mathbf{p}}\right)  &=&-\frac{1}{2}\left( 
\frac{m_{1}m_{2}}{\left( \omega _{p}+m_{1}\right) \left( \Omega
_{p}+m_{2}\right) }\right) ^{1/2}\left( \frac{\left( J+m_{J}-1\right)
(J+m_{J})}{J\left( 2J-1\right) }\right) ^{1/2}  \notag \\
&&\times \left( \gamma ^{1}+i\gamma ^{2}-i\left( \sigma ^{01}+i\sigma
^{02}\right) \right) Y_{J-1}^{m_{J}-1}\left( \widehat{\mathbf{p}}\right) ,
\end{eqnarray}
when $m_{\ell }=m_{J}$%
\begin{equation}
\Gamma _{2}^{J-1}\left( \widehat{\mathbf{p}}\right) =-\left( \frac{m_{1}m_{2}%
}{\left( \omega _{p}+m_{1}\right) \left( \Omega _{p}+m_{2}\right) }\right)
^{1/2}\left( \frac{\left( J-m_{J}\right) \left( J+m_{J}\right) }{J\left(
2J-1\right) }\right) ^{1/2}\left( \gamma ^{3}-i\sigma ^{03}\right)
Y_{J-1}^{m_{J}}\left( \widehat{\mathbf{p}}\right) ,
\end{equation}
and $m_{\ell }=m_{J}+1$%
\begin{eqnarray}
\Gamma _{3}^{J-1}\left( \widehat{\mathbf{p}}\right)  &=&\frac{1}{2}\left( 
\frac{m_{1}m_{2}}{\left( \omega _{p}+m_{1}\right) \left( \Omega
_{p}+m_{2}\right) }\right) ^{1/2}\left( \frac{\left( J-m_{J}-1\right)
(J-m_{J})}{J\left( 2J-1\right) }\right) ^{1/2}  \notag \\
&&\times \left( \gamma ^{1}-i\gamma ^{2}-i\left( \sigma ^{01}-i\sigma
^{02}\right) \right) Y_{J-1}^{m_{J}+1}\left( \widehat{\mathbf{p}}\right) .
\end{eqnarray}
\vskip0.1truecm

{\normalsize \noindent }\textbf{Pure Triplet States} $\ell =J+1,\;\ J\geq
0,\;\;P=(-1)^{J}$

\vskip0.1truecm

When $m_{\ell }=m_{J}-1$%
\begin{eqnarray}
\Gamma _{1}^{J+1}\left( \widehat{\mathbf{p}}\right)  &=&-\frac{1}{2}\left( 
\frac{m_{1}m_{2}}{\left( \omega _{p}+m_{1}\right) \left( \Omega
_{p}+m_{2}\right) }\right) ^{1/2}\left( \frac{\left( J-m_{J}+1\right) \left(
J-m_{J}+2\right) }{\left( J+1\right) \left( 2J+3\right) }\right) ^{1/2} 
\notag \\
&&\times \left( \gamma ^{1}+i\gamma ^{2}-i\left( \sigma ^{01}+i\sigma
^{02}\right) \right) Y_{J+1}^{m_{J}-1}\left( \widehat{\mathbf{p}}\right) ,
\end{eqnarray}
when $m_{\ell }=m_{J}$%
\begin{equation}
\Gamma _{2}^{J+1}\left( \widehat{\mathbf{p}}\right) =\left( \frac{m_{1}m_{2}%
}{\left( \omega _{p}+m_{1}\right) \left( \Omega _{p}+m_{2}\right) }\right)
^{1/2}\left( \frac{\left( J-m_{J}+1\right) \left( J+m_{J}+1\right) }{\left(
J+1\right) \left( 2J+3\right) }\right) ^{1/2}\left( \gamma ^{3}-i\sigma
^{03}\right) Y_{J+1}^{m_{J}}\left( \widehat{\mathbf{p}}\right) ,
\end{equation}
and $m_{\ell }=m_{J}+1$%
\begin{eqnarray}
\Gamma _{3}^{J+1}\left( \widehat{\mathbf{p}}\right)  &=&\frac{1}{2}\left( 
\frac{m_{1}m_{2}}{\left( \omega _{p}+m_{1}\right) \left( \Omega
_{p}+m_{2}\right) }\right) ^{1/2}\left( \frac{\left( J+m_{J}+2\right) \left(
J+m_{J}+1\right) }{\left( J+1\right) \left( 2J+3\right) }\right) ^{1/2} 
\notag \\
&&\times \left( \gamma ^{1}-i\gamma ^{2}-i\left( \sigma ^{01}-i\sigma
^{02}\right) \right) Y_{J+1}^{m_{J}+1}\left( \widehat{\mathbf{p}}\right) .
\end{eqnarray}

{\normalsize \newpage }

\vskip0.8truecm {\normalsize \noindent {\textbf{\large 6. The relativistic
radial equations for two-fermion systems}}} {\normalsize \vskip0.4truecm }

It is not possible to write one universal two-fermion radial wave equation,
because the adjustable functions have different form for different states.
Thus, it was important to classify all states of the system before deriving
final radial equations. Now we return to the variational equation (47) from
which we derive the radial equations for different states. We start with a
particle-antiparticle system.

\vskip0.4truecm {\normalsize \noindent }\textbf{6.1. Two-fermion wave
equations for positronium-like systems}\vskip0.2truecm

It follows from the above analysis, that three sets of radial equations
arise for this case.

\noindent \textbf{Singlet states} $\ell =J,$ $J\geq 0$, $P=\left( -1\right)
^{J+1}$, $C=\left( -1\right) ^{J}$ the radial equation is 
\begin{equation}
\left( 2\omega _{p}-E\right) f^{J}(p)=\frac{m^{2}}{N\left( 2\pi \right) ^{3}}%
\int \frac{q^{2}dq}{\omega _{p}\omega _{q}}\mathcal{K}\left( p,q\right)
f^{J}(q),
\end{equation}
with the kernel 
\begin{equation}
\mathcal{K}\left( p,q\right) =-i\sum_{i,j=1}^{3}\sum_{s_{1}s_{2}\sigma
_{1}\sigma _{2}m_{J}}\int d\widehat{\mathbf{p}}d\widehat{\mathbf{q}}\,%
\overline{u}_{\mathbf{q}\sigma _{1}}\Gamma _{i}^{J}\left( \widehat{\mathbf{q}%
}\right) v_{-\mathbf{q}\sigma _{2}}\widetilde{\mathcal{M}}_{s_{1}s_{2}\sigma
_{1}\sigma _{2}}\left( \mathbf{p,q}\right) \overline{v}_{-\mathbf{p}%
s_{2}}\Gamma _{j}^{\prime J}\left( \widehat{\mathbf{p}}\right) u_{\mathbf{p}%
s_{1}},
\end{equation}
and the normalization factor (we assume that the radial functions\thinspace\ 
$f^{J}(p)$\ have been normalized) 
\begin{eqnarray}
N &=&\int d^{3}\widehat{\mathbf{p}}\,\sum_{i=1}^{3}\sum_{s_{1}s_{2}}%
\overline{u}_{\mathbf{p}s_{1}}\Gamma _{i}^{J}\left( \widehat{\mathbf{p}}%
\right) V_{-\mathbf{p}s_{2}}\overline{V}_{-\mathbf{p}s_{2}}\Gamma
_{i}^{\prime J}\left( \widehat{\mathbf{p}}\right) u_{\mathbf{p}s_{1}} \\
&=&\int d^{3}\widehat{\mathbf{p}}\,\sum_{i=1}^{3}Tr\left[ \Gamma
_{i}^{\prime J}\left( \widehat{\mathbf{p}}\right) \frac{\gamma ^{\alpha
}p_{\alpha }+m}{2m}\Gamma _{i}^{J}\left( \widehat{\mathbf{p}}\right) \frac{%
\gamma ^{\alpha }\widetilde{p}_{\alpha }-m}{2m}\right] ,  \notag
\end{eqnarray}
where $\Gamma _{i}^{\prime }$ is the matrix $\Gamma _{i}$\ of eq. (88), but
with complex conjugate spherical functions, $Y_{\ell }^{m_{i}\ast }\left( 
\widehat{\mathbf{p}}\right) $. The four-vector $\widetilde{p}_{\alpha }$ is
defined as $\widetilde{p}_{\alpha }=(E_{p},-\mathbf{p})$.

\noindent \textbf{Triplet states }$\ell =J,$ $J\geq 0$, $P=\left( -1\right)
^{J+1}$, $C=\left( -1\right) ^{J+1}$

For these states the radial equation formally coincides with (99), however
the form of the $\Gamma _{i}^{J}$ matrix must be taken from (90)-(92).

\noindent \textbf{Triplet states }$\ell =J\pm 1$, $P=\left( -1\right) ^{J}$, 
$C=\left( -1\right) ^{J}$

In this case the variational equation (47) leads to a system of coupled
equations for the two independent radial functions $f^{J-1}(p)\;$and$%
\;f^{J+1}(p)$: 
\begin{equation}
\left( 2\omega _{p}-E\right) \mathbb{F}\left( p\right) =\frac{m^{2}}{\left(
2\pi \right) ^{3}}\int \frac{q^{2}dq}{\omega _{p}\omega _{q}}\mathbb{K}%
\left( p,q\right) \mathbb{F}\left( q\right) ,
\end{equation}
where 
\begin{equation}
\mathbb{F}\left( p\right) =\left[ 
\begin{array}{c}
N^{J-1}f^{J-1}(p) \\ 
N^{J+1}f^{J+1}(p)
\end{array}
\right] ,
\end{equation}
and 
\begin{equation}
\mathbb{K}\left( p,q\right) =\left[ 
\begin{array}{cc}
\mathcal{K}_{11}\left( p,q\right) & \mathcal{K}_{12}\left( p,q\right) \\ 
\mathcal{K}_{21}\left( p,q\right) & \mathcal{K}_{22}\left( p,q\right)
\end{array}
\right] .
\end{equation}
The kernels $\mathcal{K}_{mn}$ are 
\begin{equation}
\mathcal{K}_{mn}\left( p,q\right) =-i\sum_{i,j=1}^{3}\underset{\sigma
_{1}\sigma _{2}s_{1}s_{2}m_{J}}{\sum }\int d\hat{\mathbf{p}}\,d\hat{\mathbf{q%
}}\,\overline{u}_{\mathbf{q}\sigma _{1}}\Gamma _{i}^{\ell _{n}}\left( \hat{%
\mathbf{q}}\right) v_{-\mathbf{q}\sigma _{2}}\widetilde{\mathcal{M}}%
_{s_{1}s_{2}\sigma _{1}\sigma _{2}}\left( \mathbf{p,q}\right) \overline{v}_{-%
\mathbf{p}s_{2}}\Gamma _{j}^{\prime \ell _{m}}\left( \hat{\mathbf{p}}\right)
u_{\mathbf{p}s_{1}},
\end{equation}

Here $\ell _{1}=J-1,\;\ell _{2}=J+1$. The $\Gamma _{i}^{\ell _{n}}$ matrices
are defined by (93)-(98). The system (102) reduces to a single equation for $%
J=0$ since $f^{J-1}(p)=0$ in that case. The normalization constants $N^{J\pm
1}$ are defined by (101) with corresponding matrices $\Gamma _{i}^{J\pm 1}$
(99)-(104).

\vskip0.4truecm {\normalsize \noindent }\textbf{6.2. Two-fermion wave
equations for muonium-like systems}\vskip0.2truecm

After completing the variational procedure we obtain the following results:

\noindent \textbf{For the states with} $\ell =J=0$, $P=-1$ the radial
equation is 
\begin{equation}
\left( \omega _{p}+\Omega _{p}-E\right) f^{J}(p)=\frac{m_{1}m_{2}}{N\left(
2\pi \right) ^{3}}\int \frac{q^{2}dq}{\sqrt{\omega _{p}\omega _{q}\Omega
_{p}\Omega _{q}}}\mathcal{K}\left( p,q\right) f^{J}(q),  \label{e57}
\end{equation}
where the kernel $\mathcal{K}\left( p,q\right) $\ is defined by 
\begin{equation}
\mathcal{K}\left( p,q\right) =-\frac{i}{4\pi }\sum_{s_{1}s_{2}\sigma
_{1}\sigma _{2}}\int d\widehat{\mathbf{p}}d\widehat{\mathbf{q}}\,\overline{u}%
_{\mathbf{q}\sigma _{1}}\gamma ^{5}\left( \widehat{\mathbf{q}}\right) v_{-%
\mathbf{q}\sigma _{2}}\widetilde{\mathcal{M}}_{s_{1}s_{2}\sigma _{1}\sigma
_{2}}\left( \mathbf{p,q}\right) \overline{v}_{-\mathbf{p}s_{2}}\gamma
^{5}\left( \widehat{\mathbf{p}}\right) u_{\mathbf{p}s_{1}}.
\end{equation}
The normalization factor is 
\begin{equation}
N=\frac{1}{4\pi }\int d^{3}\widehat{\mathbf{p}}\,Tr\left[ \gamma ^{5}\frac{%
\gamma ^{\alpha }p_{\alpha 1}+m_{1}}{2m_{1}}\gamma ^{5}\frac{\gamma ^{\alpha
}\widetilde{p}_{\alpha 2}-m_{2}}{2m_{2}}\right] .
\end{equation}
\noindent \textbf{For quasi-singlet and quasi-triplet states} $\left( J\geq
1\right) $ we have the system of two equations 
\begin{equation}
\left( \omega _{p}+\Omega _{p}-E\right) \mathbb{F}\left( p\right) =\frac{%
m_{1}m_{2}}{\left( 2\pi \right) ^{3}}\int \frac{q^{2}dq}{\sqrt{\omega
_{p}\omega _{q}\Omega _{p}\Omega _{q}}}\mathbb{K}\left( p,q\right) \mathbb{F}%
\left( q\right) ,
\end{equation}
where 
\begin{equation}
\mathbb{F}\left( p\right) =\left[ 
\begin{array}{c}
N^{\left( sg\right) }f^{\left( sg\right) J}(p) \\ 
N^{\left( tr\right) }f^{\left( tr\right) J}(p)
\end{array}
\right] ,
\end{equation}
and 
\begin{equation}
\mathbb{K}\left( p,q\right) =\left[ 
\begin{array}{cc}
\mathcal{K}_{11}\left( p,q\right) & \mathcal{K}_{12}\left( p,q\right) \\ 
\mathcal{K}_{21}\left( p,q\right) & \mathcal{K}_{22}\left( p,q\right)
\end{array}
\right] .
\end{equation}
The kernels $\mathcal{K}_{mn}\left( p,q\right) $ are 
\begin{equation}
\mathcal{K}_{mn}\left( p,q\right) =-i\underset{\sigma _{1}\sigma
_{2}s_{1}s_{2}m_{J}}{\sum }\int d^{3}\widehat{\mathbf{p}}\,d^{3}\widehat{%
\mathbf{q}}\,\,\overline{u}_{\mathbf{q}\sigma _{1}}\Gamma ^{n}\left( 
\widehat{\mathbf{q}}\right) V_{-\mathbf{q}\sigma _{2}}\widetilde{\mathcal{M}}%
_{s_{1}s_{2}\sigma _{1}\sigma _{2}}\left( \mathbf{p,q}\right) \overline{V}_{-%
\mathbf{p}s_{2}}\Gamma ^{\prime m}\left( \widehat{\mathbf{p}}\right) u_{%
\mathbf{p}s_{1}}.
\end{equation}
Here we used the following notation 
\begin{eqnarray}
\Gamma ^{1}\left( \widehat{\mathbf{p}}\right) &=&\Gamma ^{J}\left( \widehat{%
\mathbf{p}}\right) ,\;\;\Gamma ^{\prime 1}\left( \widehat{\mathbf{p}}\right)
=\Gamma ^{\prime J}\left( \widehat{\mathbf{p}}\right) , \\
\Gamma ^{2}\left( \widehat{\mathbf{p}}\right) &=&\sum_{i=1}^{3}\Gamma
_{i}^{J}\left( \widehat{\mathbf{p}}\right) ,\;\;\Gamma ^{\prime 2}\left( 
\widehat{\mathbf{p}}\right) =\sum_{i=1}^{3}\Gamma _{i}^{\prime J}\left( 
\widehat{\mathbf{p}}\right) .  \notag
\end{eqnarray}
The normalization factors are 
\begin{equation}
N^{\left( sg\right) ,\left( tr\right) }=\int d^{3}\widehat{\mathbf{p}}%
\,\sum_{i=1}^{3}Tr\left[ \Gamma _{i}^{\prime J}\left( \widehat{\mathbf{p}}%
\right) \frac{\gamma ^{\alpha }p_{\alpha 1}+m_{1}}{2m_{1}}\Gamma
_{i}^{J}\left( \widehat{\mathbf{p}}\right) \frac{\gamma ^{\alpha }\widetilde{%
p}_{\alpha 2}-m_{1}}{2m_{1}}\right] .
\end{equation}

\noindent \textbf{For the triplet states} with $\ell =J\mp 1$, we have two
independent radial functions $f^{J-1}(p)\;$and$\;f^{J+1}(p)$. Thus the
variational equation (53) leads to a system of coupled equations for $%
f^{J-1}(p)\;$and$\;f^{J+1}(p)$ 
\begin{equation}
\left( \omega _{p}+\Omega _{p}-E\right) \mathbb{F}\left( p\right) =\frac{%
m_{1}m_{2}}{\left( 2\pi \right) ^{3}}\int \frac{q^{2}dq}{\sqrt{\omega
_{p}\omega _{q}\Omega _{p}\Omega _{q}}}\mathbb{K}\left( p,q\right) \mathbb{F}%
\left( q\right) ,  \label{e61}
\end{equation}
where 
\begin{equation}
\mathbb{F}\left( p\right) =\left[ 
\begin{array}{c}
N^{J-1}f^{J-1}(p) \\ 
N^{J+1}f^{J+1}(p)
\end{array}
\right] ,  \label{e62}
\end{equation}
and 
\begin{equation}
\mathbb{K}\left( p,q\right) =\left[ 
\begin{array}{cc}
\mathcal{K}_{11}\left( p,q\right) & \mathcal{K}_{12}\left( p,q\right) \\ 
\mathcal{K}_{21}\left( p,q\right) & \mathcal{K}_{22}\left( p,q\right)
\end{array}
\right] .
\end{equation}
The kernels $\mathcal{K}_{mn}$ are 
\begin{equation}
\mathcal{K}_{mn}\left( p,q\right) =-i\sum_{i,j=1}^{3}\underset{\sigma
_{1}\sigma _{2}s_{1}s_{2}m_{J}}{\sum }\int d\hat{\mathbf{p}}\,d\hat{\mathbf{q%
}}\,\overline{u}_{\mathbf{q}\sigma _{1}}\Gamma _{i}^{\ell _{n}}\left( \hat{%
\mathbf{q}}\right) V_{-\mathbf{q}\sigma _{2}}\widetilde{\mathcal{M}}%
_{s_{1}s_{2}\sigma _{1}\sigma _{2}}\left( \mathbf{p,q}\right) \overline{V}_{-%
\mathbf{p}s_{2}}\Gamma _{j}^{\prime \ell _{m}}\left( \hat{\mathbf{p}}\right)
u_{\mathbf{p}s_{1}},
\end{equation}
where $\ell _{1}=J-1,\;\ell _{2}=J+1.$ The normalization factors are defined
by analogy with (114). The system (\ref{e61}) reduces to a single equation
for $f^{J+1}(p)$ when $J=0$, since $f^{J-1}(p)=0$ in that case.

{\normalsize \newpage }

\vskip0.8truecm {\normalsize \noindent {\textbf{\large 7. Radiative
corrections to }}}$O\left( \alpha ^{5}\right) $ {\normalsize {\textbf{\large %
for arbitrary mass ratio}}}

{\normalsize \vskip0.4truecm }

The relativistic radial equations derived above cannot be solved
analytically, so an approximation method (numerical, variational, or
perturbative) must be used. We shall use first-order perturbation theory to
calculate $O\left( \alpha ^{5}\right) $\ corrections to the energy for some
states.\ 

As mentioned above, at the one-loop level radiative corrections include
vacuum polarization, fermion self-energy and vertex corrections. We
calculate their contribution to the energy shift of singlet states with $%
\ell =J=0$. These states are described by the radial equations (99) and
(106) for positronium- and muonium-like systems respectively. Since the
solution of (106) reduces to that of (99) for two equal masses, we
concentrate on equation (106) only.\ The energy eigenvalues $E_{n,J}$ can be
calculated from the equation 
\begin{eqnarray}
E_{n,J}\int_{0}^{\infty }dp\,p^{2}f^{J}(p)f^{J}(p) &=&\int_{0}^{\infty
}dp\,p^{2}\,\left( \omega _{p}+\Omega _{p}\right) f^{J}(p)f^{J}(p)  \notag \\
&&-\frac{m_{1}m_{2}}{N\left( 2\pi \right) ^{3}}\int_{0}^{\infty }\frac{%
p^{2}q^{2}dpdq}{\sqrt{\omega _{p}\Omega _{p}\omega _{q}\Omega _{q}}}\mathcal{%
K}(p\mathbf{,}q)f^{J}(p)f^{J}(q),
\end{eqnarray}
which follows from (106). For radiative corrections to $O\left( \alpha
^{5}\right) $ we take $f^{J}(p)$\ to be the non-relativistic hydrogen wave
functions in (119) and\ obtain the result 
\begin{eqnarray}
\Delta {\large \varepsilon }\left( \alpha ^{5}\right) &=&E_{n,J}-\left(
m_{1}+m_{2}\right) +\frac{\alpha ^{2}m_{r}}{2n^{2}}-\Delta {\large %
\varepsilon }\left( \alpha ^{4}\right) \\
&=&-\frac{m_{1}m_{2}}{N\left( 2\pi \right) ^{3}}\sum_{i=1,2}\int_{0}^{\infty
}\frac{p^{2}q^{2}dpdq}{\sqrt{\omega _{p}\Omega _{p}\omega _{q}\Omega _{q}}}%
\left( \mathcal{K}^{ver_{i}}(p\mathbf{,}q)+\mathcal{K}^{vac_{i}}(p\mathbf{,}%
q)+\mathcal{K}^{mass_{i}}(p\mathbf{,}q)\right) f^{J}(p)f^{J}(q),  \notag
\end{eqnarray}
where $m_{r}=m_{1}m_{2}/\left( m_{1}+m_{2}\right) $ is the reduced mass, $%
\Delta \varepsilon \left( \alpha ^{4}\right) $ are energy corrections to $%
O\left( \alpha ^{4}\right) $ derived in [2]. The kernels in (120) correspond
to the radially reduced form (107) of the second order matrix elements
(53)-(56). They are, explicitly ($\mu _{\psi }=\alpha /4\pi m_{1}$): 
\begin{eqnarray}
&&\mathcal{K}^{ver_{1}}\left( p,q\right) =\frac{Q_{1}Q_{2}\mu _{\psi }}{%
64\pi m_{1}^{2}m_{2}^{2}}\int d\hat{\mathbf{p}}\,d\hat{\mathbf{q}}\,D_{\mu
\nu }\left( p-q\right) \\
&&\times Tr\left[ \left( \gamma ^{\lambda }p_{1\lambda }+m_{1}\right) 
\mathbf{\Lambda }^{\mu }\left( p_{1}-q_{1}\right) \left( \gamma ^{\lambda
}q_{1\lambda }+m_{1}\right) \gamma ^{5}\left( \gamma ^{\lambda }\widetilde{q}%
_{2\lambda }-m_{2}\right) \gamma ^{\nu }\left( \gamma ^{\lambda }\widetilde{p%
}_{2\lambda }-m_{2}\right) \gamma ^{5}\right] ,  \notag
\end{eqnarray}
\begin{eqnarray}
&&\mathcal{K}^{vac_{1}}\left( p,q\right) =\frac{Q_{1}Q_{2}\mu _{\psi }}{%
64\pi m_{1}^{2}m_{2}^{2}}\int d\hat{\mathbf{p}}\,d\hat{\mathbf{q}}\,D_{\mu
\nu }\left( p-q\right) \mathbf{\Pi }^{\nu \alpha }\left( p_{1}-q_{1}\right)
\\
&&\times Tr\left[ \left( \gamma ^{\lambda }p_{1\lambda }+m_{1}\right) \gamma
^{\mu }\left( \gamma ^{\lambda }q_{1\lambda }+m_{1}\right) \gamma ^{5}\left(
\gamma ^{\lambda }\widetilde{q}_{2\lambda }-m_{2}\right) \gamma ^{\upsilon
}\left( \gamma ^{\lambda }\widetilde{p}_{2\lambda }-m_{2}\right) \gamma ^{5}%
\right] ,  \notag
\end{eqnarray}
\begin{eqnarray}
&&\mathcal{K}^{mass_{1}^{\prime }}\left( p,q\right) =\frac{Q_{1}Q_{2}\mu
_{\psi }}{64\pi m_{1}^{2}m_{2}^{2}}\int d\hat{\mathbf{p}}\,d\hat{\mathbf{q}}%
\,D_{\mu \nu }\left( p-q\right) \\
&&\times Tr\left[ \left( \gamma ^{\lambda }p_{1\lambda }+m_{1}\right) 
\mathbf{\Sigma }\left( p_{1}\right) S_{\psi }\left( p_{1}\right) \gamma
^{\mu }\left( \gamma ^{\lambda }q_{1\lambda }+m_{1}\right) \gamma ^{5}\left(
\gamma ^{\lambda }\widetilde{q}_{2\lambda }-m_{2}\right) \gamma ^{\nu
}\left( \gamma ^{\lambda }\widetilde{p}_{2\lambda }-m_{2}\right) \gamma ^{5}%
\right] .  \notag
\end{eqnarray}
The contribution of the kernels $\mathcal{K}^{ver_{2}}$, $\mathcal{K}%
^{vac_{2}}$ is obvious, and follows from the symmetry with respect to $m_{1}$%
\ and $m_{2}$. As will be shown, the contribution of $\mathcal{K}%
^{mass_{1}^{\prime }}$, $\mathcal{K}^{mass_{1}^{\prime \prime }}$, $\mathcal{%
K}^{mass_{2}^{\prime }}$, and $\mathcal{K}^{mass_{2}^{\prime \prime }}$\ is
zero. The renormalization of the functions $\mathbf{\Lambda }^{\mu }\left(
p_{1}-q_{1}\right) $ (64), $\mathbf{\Pi }^{\nu \alpha }\left(
p_{1}-q_{1}\right) $ (57), and $\mathbf{\Sigma }\left( p_{1}\right) $ (59)
implies that they should be replaced by the following expressions (the
justification is discussed in Ref. [16]), 
\begin{equation}
\mathbf{\Lambda }^{\mu }\left( p_{1}-q_{1}\right) \rightarrow \frac{i\alpha 
}{4\pi m_{1}}\sigma ^{\mu \nu }\left( p_{1}-q_{1}\right) _{\nu },
\end{equation}
\begin{equation}
\mathbf{\Pi }^{\nu \alpha }\left( p_{1}-q_{1}\right) \rightarrow -g^{\nu
\alpha }\frac{\alpha }{15\pi m_{1}^{2}}\left( p_{1}-q_{1}\right) ^{4},
\end{equation}
\begin{equation}
\mathbf{\Sigma }\left( p_{1}\right) \rightarrow 0.
\end{equation}
After that the calculations are straightforward and we obtain the
non-vanishing kernel contributions 
\begin{equation}
\mathcal{K}^{ver_{1}}\left( p,q\right) =-\alpha ^{2}\frac{8\pi \left(
2m_{1}-m_{2}\right) }{m_{1}^{2}m_{2}}\delta _{J,0},
\end{equation}
\begin{equation}
\mathcal{K}^{vac_{1}}\left( p,q\right) =\frac{32\pi \alpha ^{2}}{15m_{1}^{2}}%
\delta _{J,0}.
\end{equation}
To calculate energy corrections we use the nonrelativistic hydrogenic
momentum-space\ radial wave function $f^{J}(p)$ (ref. [17], eq.125). The
corrections are given by 
\begin{equation}
\Delta {\large \varepsilon }^{ver_{1}}=\frac{\alpha ^{5}m_{r}}{n^{3}}\frac{%
2\left( 2m_{1}-m_{2}\right) m_{2}}{\pi \left( m_{1}+m_{2}\right) ^{2}}\delta
_{J,0},
\end{equation}
\begin{equation}
\Delta {\large \varepsilon }^{vac_{1}}=-\frac{4\alpha ^{5}m_{r}}{15\pi n^{3}}%
\frac{m_{r}^{2}}{m_{1}^{2}}\delta _{J,0},
\end{equation}
These results agree with Ref. [18].

It should be noted that a similar treatment of positronium and muonium has
been considered by Zhang and Koniuk [16, 19]. They used postulated equations
with an inserted invariant $\widetilde{\mathcal{M}}$\ matrix. These authors
show that the inclusion of single-loop diagrams yields positronium energy
eigenstates which are accurate to $O\left( \alpha ^{5},\alpha ^{5}\ln \alpha
\right) $. In the present variational treatment the equations and results
are derived from first principles.

%\newpage
\vskip0.6truecm

{\normalsize \noindent {\textbf{\large 8. Concluding remarks}} }

{\normalsize \vskip0.4truecm }

{\normalsize %\section{Concluding remarks}
}

We{\normalsize \ }have shown that the variational method can be formulated
in a way that allows one to derive relativistic few-body equations, which
can include interactions to any order of the coupling constant. The method
is based on a reformulation of QED, in which covariant Green's functions are
used to solve partially the underlying Euler-Lagrange equations of motion.
This leads to a Hamiltonian which contains the Green's function sandwiched
between fermion currents directly (equations (14)-(16)). The eigenvalue
equation $\widehat{P}^{\mu }\,\left| \psi \right\rangle =P^{\mu }\,\left|
\psi \right\rangle $, where $\widehat{P}^{\mu }=\left( \widehat{H},\widehat{%
\mathbf{P}}\right) $\ is the energy momentum operator of the QFT is
formulated variationally, $\delta \left( \left\langle \psi \right| \widehat{H%
}-E\left| \psi \right\rangle _{t=t_{0}}\right) =0$. Time evolution operators
are used to recast the problem in such a way that Feynman diagrams of any
order can arise in the kernels that describe the interparticle interaction
(eq.(39)).

We illustrate the utility of this formulation by deriving relativistic
momentum-space wave equations for two-fermion\ systems like muonium and
positronium. These equations (47), which become (99), (102) for
particle-antiparticle systems, and (106), (109), (115) for muonium like
systems, describe the behavior of the two-fermion systems in principle to
all orders of the coupling constant for arbitrary mass ratio. For bound
states of the two-fermion system the trial states are chosen to be
eigenstates of the total angular momentum operators $\widehat{\mathbf{J}}%
^{2} $, $\widehat{J}_{3}$ and parity, and also of charge conjugation for
particle-antiparticle systems. A general relativistic reduction of the wave
equations to radial form is given for arbitrary masses of the two fermions.
For given $J$ there is a single radial equation for total spin zero singlet
states (106), but for other states there are, in general, coupled equations,
(109) for mixed-spin states, and (115) for triplet\ $\ell $-mixing states.
We have shown how the classification of the states follows naturally from
the system of eigenvalue equations (73), given our trial state (41).

We use the derived radial equations to obtain approximate perturbative
solutions for the two-fermion binding energy to $O\left( \alpha ^{5}\right) $
in the fine-structure constant for all singlet states with total angular
momentum quantum number $J=\ell =0$. Results for other states can be
obtained in an analogous manner.

The method presented here can be generalized for systems of three or more
fermions. This shall be the subject of a forthcoming work.

{\normalsize %\end{enumerate}}}
\newpage \vskip0.8truecm \noindent {\textbf{\large Acknowledgment}} }

{\normalsize \vskip0.4truecm }

{\normalsize %\section{Appendix}
%\subsection{Quantization of total angular momentum operator in relativistic
%reduction}
}

The financial support of the Natural Sciences and Engineering Research
Council of Canada is gratefully acknowledged.

{\normalsize \vskip0.8truecm \noindent {\textbf{\large Appendix A: Matrix
elements for high-order corrections}}}

{\normalsize \vskip0.4truecm }

To classify the corrections we write out the Hamiltonian density $\widehat{%
\mathcal{H}}_{Ii}\left( x\right) $ in explicit form 
\begin{equation}
\widehat{\mathcal{H}}_{Ii}\left( x\right) =\widehat{\mathcal{H}}_{I_{\psi
}}\left( x\right) +\widehat{\mathcal{H}}_{I_{\phi }}\left( x\right) +%
\widehat{\mathcal{H}}_{I_{\psi \phi }}\left( x\right) +\widehat{\mathcal{H}}%
_{I_{\phi \psi }}\left( x\right) ,
\end{equation}
where 
\begin{equation}
\widehat{\mathcal{H}}_{I_{\psi }}\left( x\right) =\frac{Q_{1}^{2}}{2}\int
d^{4}x^{\prime }\,\overline{\psi }(x^{\prime })\gamma ^{\mu }\psi (x^{\prime
})D_{\mu \nu }(x-x^{\prime })\overline{\psi }(x)\gamma ^{\nu }\psi (x),
\end{equation}
\begin{equation}
\widehat{\mathcal{H}}_{I_{\phi }}\left( x\right) =\frac{Q_{2}^{2}}{2}\int
d^{4}x^{\prime }\,\overline{\phi }(x^{\prime })\gamma ^{\mu }\phi (x^{\prime
})D_{\mu \nu }(x-x^{\prime })\overline{\phi }(x)\gamma ^{\nu }\phi (x),
\end{equation}
\begin{equation}
\widehat{\mathcal{H}}_{I_{\psi \phi }}\left( x\right) =\frac{Q_{1}Q_{2}}{2}%
\int d^{4}x^{\prime }\,\overline{\psi }(x^{\prime })\gamma ^{\mu }\psi
(x^{\prime })D_{\mu \nu }(x-x^{\prime })\overline{\phi }(x)\gamma ^{\nu
}\phi (x),
\end{equation}
\begin{equation}
\widehat{\mathcal{H}}_{I_{\phi \psi }}\left( x\right) =\frac{Q_{1}Q_{2}}{2}%
\int d^{4}x^{\prime }\,\overline{\phi }(x^{\prime })\gamma ^{\mu }\phi
(x^{\prime })D_{\mu \nu }(x-x^{\prime })\overline{\psi }(x)\gamma ^{\nu
}\psi (x),
\end{equation}

First, we consider the element $T\left( \widehat{\mathcal{H}}_{I_{\psi
}}\left( x_{1}\right) \widehat{\mathcal{H}}_{I_{\psi \phi }}\left( x\right)
\delta \left( t\right) \right) $. It is not difficult to show that 
\begin{eqnarray}
&&\left\langle \psi _{trial}\right| \,\int d^{4}x_{1}d^{4}xT\left( \widehat{%
\mathcal{H}}_{I_{\psi }}\left( x_{1}\right) \widehat{\mathcal{H}}_{I_{\psi
\phi }}\left( x\right) \delta \left( t\right) \right) \,\left| \psi
_{trial}\right\rangle \\
&=&\frac{Q_{1}^{3}Q_{2}}{4}\int d^{4}x_{1}d^{4}x_{1}^{\prime }d^{4}x^{\prime
}d^{4}xD_{\mu \nu }\left( x_{1}-x_{1}^{\prime }\right) D_{\alpha \beta
}(x-x^{\prime })\delta \left( t\right)  \notag \\
&&\times 2\left\langle \psi _{trial}\right| \left( 
\begin{array}{c}
:\overline{\psi }(x_{1}^{\prime })\gamma ^{\mu }\psi (x_{1}^{\prime
})Tr\left( iS_{\psi }\left( x^{\prime }-x_{1}\right) \gamma ^{\nu }iS_{\psi
}\left( x_{1}-x^{\prime }\right) \gamma ^{\alpha }\right) \overline{\phi }%
(x)\gamma ^{\beta }\phi (x): \\ 
+:\overline{\psi }(x_{1})\gamma ^{\nu }iS_{\psi }\left( x_{1}-x^{\prime
}\right) \gamma ^{\alpha }iS_{\psi }\left( x^{\prime }-x_{1}^{\prime
}\right) \gamma ^{\mu }\psi (x_{1}^{\prime })\overline{\phi }(x)\gamma
^{\beta }\phi (x): \\ 
+:\overline{\psi }(x_{1}^{\prime })\gamma ^{\mu }iS_{\psi }\left(
x_{1}^{\prime }-x_{1}\right) \gamma ^{\nu }iS_{\psi }\left( x_{1}-x^{\prime
}\right) \gamma ^{\alpha }\psi (x^{\prime })\overline{\phi }(x)\gamma
^{\beta }\phi (x): \\ 
+:\overline{\psi }(x^{\prime })\gamma ^{\alpha }iS_{\psi }\left( x^{\prime
}-x_{1}\right) \gamma ^{\nu }iS_{\psi }\left( x_{1}-x_{1}^{\prime }\right)
\gamma ^{\mu }\psi (x_{1}^{\prime })\overline{\phi }(x)\gamma ^{\beta }\phi
(x):
\end{array}
\right) \left| \psi _{trial}\right\rangle ,  \notag
\end{eqnarray}
where we used the symmetry of the Green's function $D_{\mu \nu }\left(
x_{1}-x_{1}^{\prime }\right) =D_{\nu \mu }\left( x_{1}^{\prime
}-x_{1}\right) $ and the standard expression for the fermion propagator
(60). Using the Fourier transform (18), (19), (49), (61) we obtain 
\begin{eqnarray}
&&\left\langle \psi _{trial}\right| \,\int d^{4}x_{1}d^{4}x\,T\left( 
\widehat{\mathcal{H}}_{I_{\psi }}\left( x_{1}\right) \widehat{\mathcal{H}}%
_{I_{\psi \phi }}\left( x\right) \delta \left( t\right) \right) \,\left|
\psi _{trial}\right\rangle \\
&=&-\frac{Q_{1}^{3}Q_{2}}{2}\frac{m_{1}m_{2}}{\left( 2\pi \right) ^{6}}%
\sum_{s_{1}^{\prime }s_{2}^{\prime }s_{1}s_{2}}\int \frac{d^{3}\mathbf{p}%
_{1}d^{3}\mathbf{p}_{2}d^{3}\mathbf{p}_{1}^{\prime }d^{3}\mathbf{p}%
_{2}^{\prime }}{\left( \omega _{p_{1}}\omega _{p_{1}^{\prime }}\Omega
_{p_{2}}\Omega _{p_{2}^{\prime }}\right) ^{1/2}}\int \frac{d^{4}k}{2\pi }%
\times  \notag \\
&&\times F_{s_{1}^{\prime }s_{2}^{\prime }}^{\ast }\left( \mathbf{p}%
_{1}^{\prime },\mathbf{p}_{2}^{\prime }\right) F_{s_{1}s_{2}}\left( \mathbf{p%
}_{1},\mathbf{p}_{2}\right) \delta \left( \mathbf{p}_{1}^{\prime }-\mathbf{p}%
_{1}+\mathbf{p}_{2}-\mathbf{p}_{2}^{\prime }\right) \times  \notag \\
&&\times \left( 
\begin{array}{c}
D_{\mu \nu }\left( p_{1}^{\prime }-p_{1}\right) D_{\alpha \beta }\left(
p_{1}^{\prime }-p_{1}\right) \times \\ 
\times \overline{u}\left( \mathbf{p}_{1}^{\prime },s_{1}^{\prime }\right)
\gamma ^{\mu }u\left( \mathbf{p}_{1},s_{1}\right) Tr\left[ S_{\psi }\left(
k-p_{1}^{\prime }\right) \gamma ^{\nu }S_{\psi }\left( k-p_{1}\right) \gamma
^{\alpha }\right] \overline{V}\left( \mathbf{p}_{2},s_{2}\right) \gamma
^{\beta }V\left( \mathbf{p}_{2}^{\prime },s_{2}^{\prime }\right) \times \\ 
+D_{\mu \nu }\left( k\right) D_{\alpha \beta }\left( p_{1}^{\prime
}-p_{1}\right) \times \\ 
\times \left( 
\begin{array}{c}
\overline{u}\left( \mathbf{p}_{1}^{\prime },s_{1}^{\prime }\right) \gamma
^{\nu }S_{\psi }\left( k+p_{1}^{\prime }\right) \gamma ^{\alpha }S_{\psi
}\left( k+p_{1}\right) \gamma ^{\mu }u\left( \mathbf{p}_{1},s_{1}\right) 
\overline{V}\left( \mathbf{p}_{2},s_{2}\right) \gamma ^{\beta }V\left( 
\mathbf{p}_{2}^{\prime },s_{2}^{\prime }\right) \\ 
+\overline{u}\left( \mathbf{p}_{1}^{\prime },s_{1}^{\prime }\right) \gamma
^{\mu }S_{\psi }\left( p_{1}^{\prime }-k\right) \gamma ^{\nu }S_{\psi
}\left( p_{1}^{\prime }\right) \gamma ^{\alpha }u\left( \mathbf{p}%
_{1},s_{1}\right) \overline{V}\left( \mathbf{p}_{2},s_{2}\right) \gamma
^{\beta }V\left( \mathbf{p}_{2}^{\prime },s_{2}^{\prime }\right) \\ 
+\overline{u}\left( \mathbf{p}_{1}^{\prime },s_{1}^{\prime }\right) \gamma
^{\alpha }S_{\psi }\left( p_{1}\right) \gamma ^{\nu }S_{\psi }\left(
k+p_{1}\right) \gamma ^{\mu }u\left( \mathbf{p}_{1},s_{1}\right) \overline{V}%
\left( \mathbf{p}_{2},s_{2}\right) \gamma ^{\beta }V\left( \mathbf{p}%
_{2}^{\prime },s_{2}^{\prime }\right)
\end{array}
\right)
\end{array}
\right)  \notag
\end{eqnarray}
The calculation of matrix elements is presented here in an arbitrary frame.
In the rest frame the adjustable functions $F_{s_{1}s_{2}}\left( \mathbf{p}%
_{1},\mathbf{p}_{2}\right) $ reduce to $F_{s_{1}s_{2}}\left( \mathbf{p}%
_{1}\right) \delta \left( \mathbf{p}_{1}+\mathbf{p}_{2}\right) $.

The next term $\widehat{\mathcal{H}}_{I_{\psi }}\left( x_{1}\right) \widehat{%
\mathcal{H}}_{I_{\phi \psi }}\left( x\right) $ yields a similar result, but
with the Green's function $D_{\alpha \beta }\left( p_{1}^{\prime
}-p_{1}\right) $ replaced by $D_{\alpha \beta }\left( p_{2}^{\prime
}-p_{2}\right) $. Finally, the terms $\widehat{\mathcal{H}}_{I_{\psi
}}\left( x_{1}\right) \widehat{\mathcal{H}}_{I_{\psi \phi }}\left( x\right) +%
\widehat{\mathcal{H}}_{I_{\psi }}\left( x_{1}\right) \widehat{\mathcal{H}}%
_{I_{\phi \psi }}\left( x\right) $\ and $\widehat{\mathcal{H}}_{I_{\psi \phi
}}\left( x_{1}\right) \widehat{\mathcal{H}}_{I_{\psi }}\left( x\right) +%
\widehat{\mathcal{H}}_{I_{\phi \psi }}\left( x_{1}\right) \widehat{\mathcal{H%
}}_{I_{\phi }}\left( x\right) $\ give the same contribution and can be
combined together in the form 
\begin{eqnarray}
&&\left\langle \psi _{trial}\right| \,\int d^{4}x_{1}d^{4}x\,T\left( \left( 
\begin{array}{c}
\widehat{\mathcal{H}}_{I_{\psi }}\left( x_{1}\right) \widehat{\mathcal{H}}%
_{I_{\psi \phi }}\left( x\right) +\widehat{\mathcal{H}}_{I_{\psi }}\left(
x_{1}\right) \widehat{\mathcal{H}}_{I_{\phi \psi }}\left( x\right) \\ 
+\widehat{\mathcal{H}}_{I_{\psi \phi }}\left( x_{1}\right) \widehat{\mathcal{%
H}}_{I_{\psi }}\left( x\right) +\widehat{\mathcal{H}}_{I_{\phi \psi }}\left(
x_{1}\right) \widehat{\mathcal{H}}_{I_{\phi }}\left( x\right)
\end{array}
\right) \delta \left( t\right) \right) \,\left| \psi _{trial}\right\rangle 
\notag \\
&=&-\frac{m_{1}m_{2}}{\left( 2\pi \right) ^{3}}\sum_{s_{1}^{\prime
}s_{2}^{\prime }s_{1}s_{2}}\int \frac{d^{3}\mathbf{p}_{1}d^{3}\mathbf{p}%
_{2}d^{3}\mathbf{p}_{1}^{\prime }d^{3}\mathbf{p}_{2}^{\prime }}{\left(
\omega _{p_{1}}\omega _{p_{1}^{\prime }}\Omega _{p_{2}}\Omega
_{p_{2}^{\prime }}\right) ^{1/2}}F_{s_{1}^{\prime }s_{2}^{\prime }}^{\ast
}\left( \mathbf{p}_{1}^{\prime },\mathbf{p}_{2}^{\prime }\right)
F_{s_{1}s_{2}}\left( \mathbf{p}_{1},\mathbf{p}_{2}\right) \delta \left( 
\mathbf{p}_{1}^{\prime }-\mathbf{p}_{1}+\mathbf{p}_{2}-\mathbf{p}%
_{2}^{\prime }\right) \left( -i\right) \times  \notag \\
&&\times \left( 
\begin{array}{c}
\mathcal{M}_{s_{1}^{\prime }s_{2}^{\prime }s_{1}s_{2}}^{\left( 2\right)
vac_{1}}\left( \mathbf{p}_{1},\mathbf{p}_{1}^{\prime },\mathbf{p}_{2},%
\mathbf{p}_{2}^{\prime }\right) +\mathcal{M}_{s_{1}^{\prime }s_{2}^{\prime
}s_{1}s_{2}}^{\left( 2\right) ver_{1}}\left( \mathbf{p}_{1},\mathbf{p}%
_{1}^{\prime },\mathbf{p}_{2},\mathbf{p}_{2}^{\prime }\right) \\ 
+\mathcal{M}_{s_{1}^{\prime }s_{2}^{\prime }s_{1}s_{2}}^{\left( 2\right)
mass_{1}^{\prime }}\left( \mathbf{p}_{1},\mathbf{p}_{1}^{\prime },\mathbf{p}%
_{2},\mathbf{p}_{2}^{\prime }\right) +\mathcal{M}_{s_{1}^{\prime
}s_{2}^{\prime }s_{1}s_{2}}^{\left( 2\right) mass_{1}^{\prime \prime
}}\left( \mathbf{p}_{1},\mathbf{p}_{1}^{\prime },\mathbf{p}_{2},\mathbf{p}%
_{2}^{\prime }\right)
\end{array}
\right) ,
\end{eqnarray}

where $\mathcal{M}_{s_{1}^{\prime }s_{2}^{\prime }s_{1}s_{2}}^{\left(
2\right) vac_{1}}$, $\mathcal{M}_{s_{1}^{\prime }s_{2}^{\prime
}s_{1}s_{2}}^{\left( 2\right) ver_{1}}$,\ $\mathcal{M}_{s_{1}^{\prime
}s_{2}^{\prime }s_{1}s_{2}}^{\left( 2\right) mass_{1}^{\prime }}$, $\mathcal{%
M}_{s_{1}^{\prime }s_{2}^{\prime }s_{1}s_{2}}^{\left( 2\right)
mass_{1}^{\prime \prime }}$ are defined by (53)-(56).

Similarly we can show that the next terms contribute to the two-photon
exchange process 
\begin{eqnarray*}
&&\left\langle \psi _{trial}\right| \,\int d^{4}x_{1}d^{4}x\,T\left( \left( 
\begin{array}{c}
\widehat{\mathcal{H}}_{I_{\psi \phi }}\left( x_{1}\right) \widehat{\mathcal{H%
}}_{I_{\psi \phi }}\left( x\right) +\widehat{\mathcal{H}}_{I_{\psi \phi
}}\left( x_{1}\right) \widehat{\mathcal{H}}_{I_{\phi \psi }}\left( x\right)
\\ 
+\widehat{\mathcal{H}}_{I_{\phi \psi }}\left( x_{1}\right) \widehat{\mathcal{%
H}}_{I_{\psi \phi }}\left( x\right) +\widehat{\mathcal{H}}_{I_{\phi \psi
}}\left( x_{1}\right) \widehat{\mathcal{H}}_{I_{\phi \psi }}\left( x\right)
\end{array}
\right) \delta \left( t\right) \right) \,\left| \psi _{trial}\right\rangle \\
&=&-\frac{q_{1}q_{2}m_{1}m_{2}}{2\left( 2\pi \right) ^{6}}%
\sum_{s_{1}^{\prime }s_{2}^{\prime }s_{1}s_{2}}\int \frac{d^{3}\mathbf{p}%
_{1}d^{3}\mathbf{p}_{2}d^{3}\mathbf{p}_{1}^{\prime }d^{3}\mathbf{p}%
_{2}^{\prime }}{\left( \omega _{p_{1}^{\prime }}\omega _{p_{1}}\Omega
_{p_{2}}\Omega _{p_{2}^{\prime }}\right) ^{1/2}}\delta \left( \mathbf{p}_{2}-%
\mathbf{p}_{1}+\mathbf{p}_{1}^{\prime }-\mathbf{p}_{2}^{\prime }\right)
\times \\
&&\times F_{s_{1}^{\prime }s_{2}^{\prime }}^{\ast }\left( \mathbf{p}%
_{1}^{\prime },\mathbf{p}_{2}^{\prime }\right) F_{s_{1}s_{2}}\left( \mathbf{p%
}_{1},\mathbf{p}_{2}\right) \left( -i\right) \mathcal{M}_{s_{1}^{\prime
}s_{2}^{\prime }s_{1}s_{2}}^{2\gamma }\left( \mathbf{p}_{1},\mathbf{p}%
_{1}^{\prime },\mathbf{p}_{2},\mathbf{p}_{2}^{\prime }\right) ,
\end{eqnarray*}
where $\mathcal{M}_{s_{1}^{\prime }s_{2}^{\prime }s_{1}s_{2}}^{2\gamma }$ is
defined by (63)-(64).

{\normalsize \vskip2.8truecm \noindent {\textbf{\large Appendix B: The
one-loop renormalization scheme}}}

{\normalsize \vskip0.4truecm }

It is well known that the vacuum polarization function (57) can be written
as 
\begin{equation}
\mathbf{\Pi }^{\nu \alpha }\left( p_{1}^{\prime }-p_{1}\right) =\left[
\left( p_{1}^{\prime }-p_{1}\right) ^{\nu }\left( p_{1}^{\prime
}-p_{1}\right) ^{\alpha }-\left( p_{1}^{\prime }-p_{1}\right) ^{2}g^{\nu
\alpha }\right] \,\mathbf{\Pi }\left( p_{1}^{\prime }-p_{1}\right) ,
\end{equation}
where 
\begin{equation}
\mathbf{\Pi }\left( p_{1}^{\prime }-p_{1}\right) =\Pi \left( 0\right) +%
\overline{\Pi }\left( p_{1}^{\prime }-p_{1}\right) ,
\end{equation}
\begin{eqnarray}
\Pi _{1}\left( 0\right) &=&-\frac{iQ_{1}^{2}}{12\pi ^{4}}\int \frac{d^{4}k}{%
\left( k^{2}-m_{1}+i\varepsilon \right) ^{2}}, \\
\overline{\Pi }\left( p_{1}^{\prime }-p_{1}\right) &=&-\frac{Q_{1}^{2}}{2\pi
^{2}}\int_{0}^{1}dzz\left( 1-z\right) \ln \left[ 1-\frac{\left(
p_{1}^{\prime }-p_{1}\right) ^{2}z\left( 1-z\right) }{m_{1}^{2}}\right] .
\end{eqnarray}
The infinite sum 
\begin{eqnarray}
&&iD_{\mu \beta }^{\prime }\left( p_{1}^{\prime }-p_{1},p_{2}^{\prime
}-p_{2}\right) \\
&=&iD_{\mu \beta }\left( p_{1}^{\prime }-p_{1},p_{2}^{\prime }-p_{2}\right)
+D_{\mu \nu }\left( p_{1}^{\prime }-p_{1}\right) \mathbf{\Pi }^{\nu \alpha
}\left( p_{1}^{\prime }-p_{1}\right) iD_{\alpha \beta }\left( p_{1}^{\prime
}-p_{1},p_{2}^{\prime }-p_{2}\right)  \notag \\
&&+D_{\mu \nu }\left( p_{1}^{\prime }-p_{1}\right) \mathbf{\Pi }^{\nu \alpha
}\left( p_{1}^{\prime }-p_{1}\right) D_{\alpha \delta }\left( p_{1}^{\prime
}-p_{1}\right) \mathbf{\Pi }^{\delta \sigma }\left( p_{1}^{\prime
}-p_{1}\right) iD_{\sigma \beta }\left( p_{1}^{\prime }-p_{1},p_{2}^{\prime
}-p_{2}\right)  \notag \\
&&+...,  \notag
\end{eqnarray}
gives the dressed Green's function, if the vacuum polarization is due to the
first particle. The bare Green's function $D_{\mu \beta }\left(
p_{1}^{\prime }-p_{1},p_{2}^{\prime }-p_{2}\right) $\ is defined by (50)
and, in Feynman gauge, has the\ form 
\begin{equation}
D_{\mu \beta }\left( p_{1}^{\prime }-p_{1},p_{2}^{\prime }-p_{2}\right)
=g_{\mu \beta }\left( \frac{1}{-\left( p_{1}^{\prime }-p_{1}\right) ^{2}}+%
\frac{1}{-\left( p_{2}^{\prime }-p_{2}\right) ^{2}}\right) .
\end{equation}
Dropping terms like $\left( p_{1}^{\prime }-p_{1}\right) ^{\nu }\left(
p_{1}^{\prime }-p_{1}\right) ^{\alpha }$ (they contribute zero when they are
contracted into the final or initial currents) we obtain 
\begin{eqnarray}
&&iD_{\mu \beta }^{\prime }\left( p_{1}^{\prime }-p_{1},p_{2}^{\prime
}-p_{2}\right) \\
&\Rightarrow &\left( 1-\mathbf{\Pi }\left( p_{1}^{\prime }-p_{1}\right) +%
\mathbf{\Pi }^{2}\left( p_{1}^{\prime }-p_{1}\right) -\mathbf{\Pi }%
^{3}\left( p_{1}^{\prime }-p_{1}\right) +..\right) iD_{\mu \beta }\left(
p_{1}^{\prime }-p_{1},p_{2}^{\prime }-p_{2}\right)  \notag \\
&=&\frac{1}{1+\mathbf{\Pi }\left( p_{1}^{\prime }-p_{1}\right) }iD_{\mu
\beta }\left( p_{1}^{\prime }-p_{1},p_{2}^{\prime }-p_{2}\right)  \notag
\end{eqnarray}
If we take into account polarization due to the second particle we obtain 
\begin{eqnarray}
&&iD_{\mu \beta }^{\prime }\left( p_{1}^{\prime }-p_{1},p_{2}^{\prime
}-p_{2}\right) \\
&=&\frac{1}{1+\mathbf{\Pi }\left( p_{1}^{\prime }-p_{1}\right) }iD_{\mu
\beta }\left( p_{1}^{\prime }-p_{1},p_{2}^{\prime }-p_{2}\right) \frac{1}{1+%
\mathbf{\Pi }\left( p_{2}^{\prime }-p_{2}\right) }  \notag \\
&=&\frac{1}{1+\Pi _{1}\left( 0\right) +\overline{\mathbf{\Pi }}\left(
p_{1}^{\prime }-p_{1}\right) }iD_{\mu \beta }\left( p_{1}^{\prime
}-p_{1},p_{2}^{\prime }-p_{2}\right) \frac{1}{1+\Pi _{2}\left( 0\right) +%
\overline{\mathbf{\Pi }}\left( p_{2}^{\prime }-p_{2}\right) }  \notag
\end{eqnarray}
Then we define 
\begin{equation}
\frac{1}{Z_{3}^{\left( 1\right) }}=1+\Pi _{1}\left( 0\right) ,\;\;\;\frac{1}{%
Z_{3}^{\left( 2\right) }}=1+\Pi _{2}\left( 0\right) ,
\end{equation}
and obtain 
\begin{equation}
iD_{\mu \beta }^{\prime }\left( p_{1}^{\prime }-p_{1},p_{2}^{\prime
}-p_{2}\right) =\frac{Z_{3}^{\left( 1\right) }}{1+Z_{3}^{\left( 1\right) }%
\overline{\mathbf{\Pi }}\left( p_{1}^{\prime }-p_{1}\right) }iD_{\mu \beta
}\left( p_{1}^{\prime }-p_{1},p_{2}^{\prime }-p_{2}\right) \frac{%
Z_{3}^{\left( 2\right) }}{1+Z_{3}^{\left( 2\right) }\overline{\mathbf{\Pi }}%
\left( p_{2}^{\prime }-p_{2}\right) }.
\end{equation}
Introducing the renormalized quantities 
\begin{eqnarray}
\mathbf{\Pi }_{R}\left( p_{1}^{\prime }-p_{1}\right) &=&Z_{3}^{\left(
1\right) }\overline{\mathbf{\Pi }}\left( p_{1}^{\prime }-p_{1}\right) , \\
\mathbf{\Pi }_{R}\left( p_{2}^{\prime }-p_{2}\right) &=&Z_{3}^{\left(
2\right) }\overline{\mathbf{\Pi }}\left( p_{2}^{\prime }-p_{2}\right) , 
\notag
\end{eqnarray}
expression (160) can be written as 
\begin{equation}
iD_{\mu \beta }^{\prime }\left( p_{1}^{\prime }-p_{1},p_{2}^{\prime
}-p_{2}\right) =\frac{Z_{3}^{\left( 1\right) }Z_{3}^{\left( 2\right) }}{%
\left( 1+\mathbf{\Pi }_{R}\left( p_{1}^{\prime }-p_{1}\right) \right) \left(
1+\mathbf{\Pi }_{R}\left( p_{2}^{\prime }-p_{2}\right) \right) }iD_{\mu
\beta }\left( p_{1}^{\prime }-p_{1},p_{2}^{\prime }-p_{2}\right)
\end{equation}
Constants $Z_{3}^{\left( 1\right) }$\ and $Z_{3}^{\left( 2\right) }$ shall
be absorbed by renormalization of charges. In the lowest order of expansion,
the dressed Green's function (150) is 
\begin{equation}
iD_{\mu \beta }^{\prime }\left( p_{1}^{\prime }-p_{1},p_{2}^{\prime
}-p_{2}\right) =Z_{3}^{\left( 1\right) }Z_{3}^{\left( 2\right) }\left( 1-%
\mathbf{\Pi }_{R}\left( p_{1}^{\prime }-p_{1}\right) -\mathbf{\Pi }%
_{R}\left( p_{2}^{\prime }-p_{2}\right) \right) iD_{\mu \beta }\left(
p_{1}^{\prime }-p_{1},p_{2}^{\prime }-p_{2}\right)
\end{equation}
Note, that in the nonrelativistic limit the $\mathbf{\Pi }_{R}$- functions
take on the form 
\begin{equation}
\mathbf{\Pi }_{R}\left( p_{1}^{\prime }-p_{1}\right) =\frac{\alpha \left(
p_{1}^{\prime }-p_{1}\right) ^{2}}{15\pi m_{1}^{2}},\;\;\;\mathbf{\Pi }%
_{R}\left( p_{2}^{\prime }-p_{2}\right) =\frac{\alpha \left( p_{2}^{\prime
}-p_{2}\right) ^{2}}{15\pi m_{2}^{2}}.
\end{equation}
Next, we consider the vertex function 
\begin{equation}
\mathbf{\Gamma }^{\alpha }\left( p_{1}^{\prime },p_{1}\right) =\gamma
^{\alpha }+\mathbf{\Lambda }^{\alpha }\left( p_{1}^{\prime },p_{1}\right) ,
\end{equation}
where $\mathbf{\Lambda }^{\alpha }$\ can be written in the form\ 
\begin{equation}
\mathbf{\Lambda }^{\alpha }\left( p_{1}^{\prime },p_{1}\right) =\mathcal{F}%
_{1}\left( p_{1}^{\prime },p_{1}\right) \gamma ^{\alpha }+\mathcal{F}%
_{2}\left( p_{1}^{\prime },p_{1}\right) \frac{i\sigma ^{\alpha \beta }\left(
p_{1}^{\prime }-p_{1}\right) _{\beta }}{2m_{1}},
\end{equation}
where the scalar function $\mathcal{F}_{1}\left( p_{1}^{\prime
},p_{1}\right) $ diverges, while $\mathcal{F}_{2}\left( p_{1}^{\prime
},p_{1}\right) $ is finite. Therefor, 
\begin{eqnarray}
&&\mathbf{\Gamma }^{\alpha }\left( p_{1}^{\prime },p_{1}\right) \\
&=&\gamma ^{\alpha }+\mathcal{F}_{1}\left( p_{1}^{\prime },p_{1}\right)
\gamma ^{\alpha }+\mathcal{F}_{2}\left( p_{1}^{\prime },p_{1}\right) \frac{%
i\sigma ^{\alpha \beta }\left( p_{1}^{\prime }-p_{1}\right) _{\beta }}{2m_{1}%
}  \notag \\
&=&\left( 1+\mathcal{F}_{1}\left( p_{1}^{\prime },p_{1}^{\prime }\right)
\right) \gamma ^{\alpha }+\left( \mathcal{F}_{1}\left( p_{1}^{\prime
},p_{1}\right) -\mathcal{F}_{1}\left( p_{1}^{\prime },p_{1}^{\prime }\right)
\right) \gamma ^{\alpha }+\mathcal{F}_{2}\left( p_{1}^{\prime },p_{1}\right) 
\frac{i\sigma ^{\alpha \beta }\left( p_{1}^{\prime }-p_{1}\right) _{\beta }}{%
2m_{1}}.  \notag
\end{eqnarray}
Then we define a value 
\begin{equation}
\frac{1}{Z_{1}}=1+\mathcal{F}_{1}\left( p_{1}^{\prime },p_{1}^{\prime
}\right) ,
\end{equation}
which will be absorbed by the renormalized charge 
\begin{equation}
Q_{1R}=Q_{1}\frac{1}{Z_{1}}
\end{equation}
The renormalization of the self-energy can be treated similarly.

{\normalsize %\newpage
\vskip0.8truecm }

{\normalsize \noindent {\textbf{\large References}} \vskip0.4truecm }

{\normalsize \enumerate}

1. A. G. Terekidi, J. W. Darewych, Journal of Mathematical Physics \textbf{45%
}, 1474 (2004).

2. A. G. Terekidi, J. W. Darewych, Journal of Mathematical Physics \textbf{46%
}, 032302 (2005).

3. J. W. Darewych, Annales Fond. L. de Broglie (Paris) 23, 15 (1998).

4. J. W. Darewych, in \textit{Causality and Locality in Modern Physics}, G
Hunter et al. (eds.), p. 333, (Kluwer, 1998).

5. J. W. Darewych, Can. J. Phys. \textbf{76}, 523 (1998).

6. M. Barham and J. W. Darewych, J. Phys. A \textbf{31}, 3481 (1998).

7. B. Ding and J. Darewych, J. Phys. G \textbf{26}, 907 (2000).

8. J. D. Jackson, \textit{Classical Electrodynamics} (John Wiley, New York,
1975).

9. A. O. Barut, \textit{Electrodynamics and Classical Theory of Fields and
Particles} (Dover, New York, 1980).

10. W. T. Grandy, \textit{Relativistic Quantum Mechanics of Leptons and
Fields} (Kluwer, 1991).

11. A. O. Barut, in \textit{Geometrical and Algebraic Aspects of Nonlinear
Field Theory}, edited by S. De Filippo, M. Marinaro, G. Marmo and G. Vilasi,
(Elsevier New York, 1989), p. 37.

12. M. Gell-Mann, F. Low, Phys. Rev. \textbf{84}, 350 (1951)

13. A. Fetter, J. Walecka, \textit{Quantum Theory of Many-Particle Systems}
(McGrow-Hill, 1971).

14. J. W. Darewych and M. Horbatsch, J. Phys. B: At. Mol. Opt. \textbf{22},
973 (1989); \textbf{23}, 337 (1990).

15. W. Dykshoorn and R. Koniuk, Phys. Rev. A \textbf{41}, 64 (1990).

16. T. Zhang and R. Koniuk, Can. J. Phys. \textbf{70}, 683 (1992).

17. H. A. Bethe and E. E. Salpeter, \textit{Quantum Mechanics of One- and
Two-Electron Atoms }(Springer, 1957).

18. T. Fulton, P. Martin Phys. Rev. \textbf{95}, 811 (1954).

19. T. Zhang and R. Koniuk, Can. J. Phys. \textbf{70}, 670 (1992).

\end{document}